\def\cleanversion{1}
\documentclass[12pt,fleqn]{iopart}

\expandafter\let\csname equation*\endcsname\relax
\expandafter\let\csname endequation*\endcsname\relax

\usepackage{amsmath}

\usepackage{iopams}

\usepackage{graphicx}
\usepackage{subcaption}
\usepackage{hyperref}
\usepackage{placeins}
\usepackage{xcolor}
\usepackage{soul}
\usepackage[normalem]{ulem}
\sethlcolor{red}

\definecolor{revisionpurple}{HTML}{7030A0}
\ifdefined\cleanversion
\newcommand{\added}[1]{#1}
\newcommand{\deleted}[1]{}

\hypersetup{hidelinks}
\else
\newcommand{\added}[1]{\textcolor{revisionpurple}{#1}}
\newcommand{\deleted}[1]{\textcolor{red}{\sout{#1}}}
\hypersetup{
     colorlinks   = true,
     citecolor    = blue,
     linkcolor    = blue}
\fi

\newcommand{\addedRtwo}[1]{#1}
\newcommand{\deletedRtwo}[1]{}

\renewcommand{\v}[1]{\ensuremath{\mathbf{#1}}} 
\newcommand{\bv}[1]{\ensuremath{\mathbf{#1}}} 
 
\newcommand{\uv}[1]{\ensuremath{\mathbf{\hat{#1}}}} 
\newcommand{\abs}[1]{\left| #1 \right|} 
\newcommand{\pd}[2]{\frac{\partial #1}{\partial #2}} 
 
\newcommand{\grad}[1]{\nabla #1} 
\renewcommand{\div}[1]{\nabla \cdot #1} 
\let\baraccent=\= 
\renewcommand{\=}[1]{\stackrel{#1}{=}} 
\newcommand{\gradperp}[1]{{{\nabla}_{\perp}} {#1}} 

\newcommand{\kperp}{k_\perp}
\newcommand{\rhos}{\rho_s}
\newcommand{\vperp}{v_\perp}
\newcommand{\vpar}{v_\parallel}

\newcommand{\dvpar}{\mathrm{d}v_\parallel}
\newcommand{\dmu}{\mathrm{d}\mu}

\newcommand{\geoJac}{J_x}
\newcommand{\phaseJac}{\mathcal{J}}
\newcommand{\cmag}{\mathcal{C}}
\newcommand{\bmag}{B}
\newcommand{\Bstarpar}{B^*_\parallel}
\newcommand{\bhat}{\uv{b}}
\newcommand{\vBstar}{\bv{B^*}}
\newcommand{\tOMP}{\mathrm{OMP}}
\newcommand{\tLCFS}{\mathrm{LCFS}}
\newcommand{\taxis}{\mathrm{axis}}
\newcommand{\Raxis}{R_{\taxis}}
\newcommand{\Baxis}{B_{\taxis}}

\newcommand{\recR}{\ensuremath{R}}

\newcommand{\upars}{u_{\parallel s}}
\newcommand{\Tpare}{T_{\parallel e}}
\newcommand{\Tpari}{T_{\parallel i}}

\newcommand{\Nz}{N_{z}}
\newcommand{\Nvpar}{N_{\vpar}}
\newcommand{\Nmu}{N_{\mu}}

\newcommand{\gke}{\textsc{Gkeyll}~}
\newcommand{\ltx}{LTX-$\beta$~}

\newcommand{\ignore}[1]{}  

\begin{document}

\title[Kinetic parallel transport in low-recycling SOL: A 1D gyrokinetic study]{Kinetic modifications to parallel transport in a low-recycling scrape-off layer: A one dimensional gyrokinetic study}

\author{A. Maan$^1$, E. Perez$^2$, M. Francisquez$^1$, M. Dorf$^3$, D. P. Boyle$^1$, R. Majeski$^1$, A. Hakim$^1$, G. W. Hammett$^1$}
\address{$^1$Princeton Plasma Physics Laboratory. Princeton, NJ. USA.}
\address{$^2$University of Wisconsin at Madison. Madison, WI. USA.}
\address{$^3$Lawrence Livermore National Laboratory, Livermore, California 94550, USA}
\ead{amaan@pppl.gov}
\vspace{10pt}

\begin{abstract}
 A liquid lithium boundary offers a possible route to improved energy confinement via reduced recycling, but the resulting low-density, high-temperature edge can become sufficiently collisionless such that standard fluid models are no longer applicable. Experimental measurements of scrape-off-layer (SOL) parameters, including density decay length ($\lambda_n$) estimates presented here, can be used for simple analytic estimates of kinetic effects, but more complete models are needed. We present a gyrokinetic simulations, with one configuration-space dimension and two velocity-space dimensions, of low and higher-recycling \ltx SOL conditions using \gke\unskip, an open-source continuum gyrokinetic code. The simulations follow a single open field line from the low-field-side midplane to the high-field-side limiter. \gke is used to evolve kinetic ion and electron distribution functions, the electrostatic potential, collisions, sources, and sheath losses with experimentally constrained density and temperature at the low field side mid-plane. The low-recycling case develops a hotter SOL, a larger electrostatic potential variation along the field line, and a density response that is not recovered with an adiabatic Boltzmann-electron closure of the field equations. This indicates that kinetic-electron dynamics and the associated ambipolar potential are important for maintaining quasi-neutrality and reducing electron end loss from the field line in the low-recycling regime. Collisionality scans show that, as the ion neoclassical collisionality is reduced below unity, the ion distribution develops a stronger non-Maxwellian structure. The corresponding SOL particle and energy confinement times increase at low collisionality, especially for the low-recycling case, because ions become mirror trapped, and electron transport is suppressed by the ambipolar electrostatic potential to maintain quasi-neutrality. These results suggest that low-recycling lithium-wall operation in \ltx accesses a SOL regime in which mirror trapping and electrostatic modification of electron transport alter the field-line residence time relative to conventional high-recycling fluid expectations. These SOL modifications are expected to become more pronounced in larger devices.
\end{abstract}

%
%
%
%
%

\section{Introduction}

Fusion pilot plants must address the divertor power-exhaust challenge due to narrow scrape-off-layer heat flux widths ($\lambda_q$) and correspondingly large parallel heat fluxes \cite{menard,menard2022,goldston2012,eich2013,kuang2020}. Under such extreme exhaust conditions, even the best solid (tungsten) plasma-facing components (PFCs) face several well-known challenges~\cite{yang2015,wright2022,matthews2013,putterichJET}. Liquid metal plasma facing surfaces/components are seen as an attractive alternative, with flow and/or evaporation offering possible ways to resiliently handle heat fluxes \cite{Fisher2020divertorlets,Goldston2016vaporbox}. 

Beyond general liquid metal effects on power handling, lithium can have a strong effect on SOL physics and heat fluxes due to its unique ability to greatly reduce the recycling of hydrogen plasma incident on the PFCs.  In steady state, solid high-Z PFCs recycle all incoming plasma and build up a neutral population that readily charge exchanges with the incoming SOL plasma and acts as an energy sink, clamping edge temperatures \cite{Groth1989,Wang2024pedestal,kaita-ppcf}. By contrast, lithium can retain up to 85$\%$ of the incoming ions \cite{krstic-li2o,baldwin2002}. \added{Hydrogen reflection and retention also depend on the composition and chemical state of lithium-containing surfaces~\cite{krstic2023detailed}.} With reduced recycling, the neutral source and associated charge-exchange cooling are strongly reduced, so the edge is expected to run hotter than in a conventional high-recycling boundary. In the limiting case where edge cooling mechanisms are removed, low-recycling theory predicts that the \deleted{edge temperature} \added{temperature at the edge of the confined plasma} can approach the core temperature, producing flat or nearly flat temperature profiles \cite{Zakharov2004LiWall,Catto2006Isomak,Smirnov2010lowrecycling}. This low-recycling, high-temperature edge regime was anticipated and observed in LTX in the electron channel, where lithium plasma-facing surfaces remained compatible with elevated edge temperatures and reduced edge collisionality \cite{Majeski2017}. Later \ltx\added{~\cite{elliott2020initial}} experiments extended this flat-temperature, low-recycling operating regime with ohmic and neutral-beam heating \cite{Boyle2023NF,krstic2025diffusion,perez2026testing,le2026synthetic,maan2026photodiode,maan2026development,maan2026plasma}.

\added{Increasing lithium wall coatings in \ltx has been associated with improved neutral and plasma density control~\cite{maan_nme_2023}; edge neutrals also affect tearing-mode activity and access to flat temperature profiles~\cite{banerjee2024investigating}. Vacuum-transfer methods enable characterization of plasma-exposed PFCs~\cite{maan2020simple}, and surface measurements in \ltx have correlated lithium PFC composition and oxidation with plasma conditions and performance~\cite{maan_ieee_2020,maan_ppcf_2021}.}

The resulting combination of high edge temperature and low density naturally pushes the SOL toward weak collisionality, where kinetic parallel dynamics and mirror trapping can become important. The kinetic parallel dynamics of a low-recycling SOL have not been explored in detail, and the implications for SOL profiles and heat fluxes are potentially very important but not yet well understood. Experimental SOL decay widths have also not previously been reported in the low-recycling regime. Experimental data from \ltx indicates that low-recycling discharges typically exhibit densities and temperatures consistent with a kinetic scrape-off layer \cite{maan2024recycling}, with ion and electron collisionalities between 1/10 and 1/100 at the last closed flux surface (LCFS). 

In this paper we present results from 1D full-$f$ gyrokinetic simulations of the scrape-off layer of \ltx using the \gke code \cite{gkeyll}. The simulations are performed along a field line that connects the low field side mid-plane to the high field side limiter. The simulations are constrained by \deleted{density and temperature measurements from a low recycling ($R\sim0.5$) and a higher recycling discharge ($R\sim0.8$, so still much lower than most other tokamaks)} \added{density and temperature measurements from representative discharges in a low-recycling ensemble ($R=0.61\text{--}0.67$) and a high-recycling ensemble ($R=0.83\text{--}0.87$) \cite{maan2024recycling}}. Because the plasma shapes of the two shots are not perfectly identical, we use the equilibrium of the high-recycling shot for a more direct comparison, allowing us to isolate how the low and high-recycling plasma regimes differ in along-the-field-line particle and energy transport. 

It is seen from collisionality scans of the open field line that the ion distribution at the low field side mid-plane has a large non-Maxwellian fraction. \ltx is a low aspect ratio tokamak with limiting surfaces at the high field side. The plasma, therefore, sees a modest mirror ratio of $\sim$ 4 from the low field side midplane to high field side limiter along the field line. At lower collisionality this effect likely gives rise to mirror trapping along the open field line, since trapped particle trajectories are not destroyed collisionally. To test this hypothesis we also simulated the same field line but with homogeneous magnetic field magnitude, effectively removing the mirror force. The simulations with constant magnetic field do not exhibit poloidal density peaking (\ref{sec:constBsims}).

The manuscript is organized as follows. The low and high-recycling discharges are described in section 2, including Thomson and probe measurements of core and SOL profiles along with estimates of SOL density decay lengths. Section 3 details the full-$f$ gyrokinetic model (one-dimensional in configuration space and two-dimensional in velocity space) implemented in \gke\unskip, describing the field-aligned coordinate geometry, the collision operator, the source terms, and two different electrostatic field models. Section 4 presents the results of these simulations, comparing the high and low-recycling cases and the different electrostatic models, followed by an ion collisionality scan to explore mirror trapping and the effect on SOL confinement times. Section 5 offers concluding remarks, using simple particle-balance arguments to discuss the simulated changes in confinement in the context of the experimentally observed density decay length trends. Finally, the appendices detail convergence studies and comparison simulations without the mirror force. 


\section{\ltx low and high-recycling discharges} \label{sec:setup}
The experiments in this work were performed in \ltx\unskip, with solid lithium coatings applied to the stainless steel plasma facing surfaces of its internal shell. The shell conforms to a limited plasma with a major radius of 0.4 m and a minor radius of 0.26 meters \cite{Majeski2017,maan2024recycling}. Two lithium evaporators, placed diametrically opposite to each other, coat $>$ 96$\%$ of the plasma-facing shell surface. The average coating thickness is greater than the maximum plasma interaction depth \cite{maan_nme_2023}. 

\begin{figure}[htbp]
  \centering
  \includegraphics[width=0.45\textwidth]{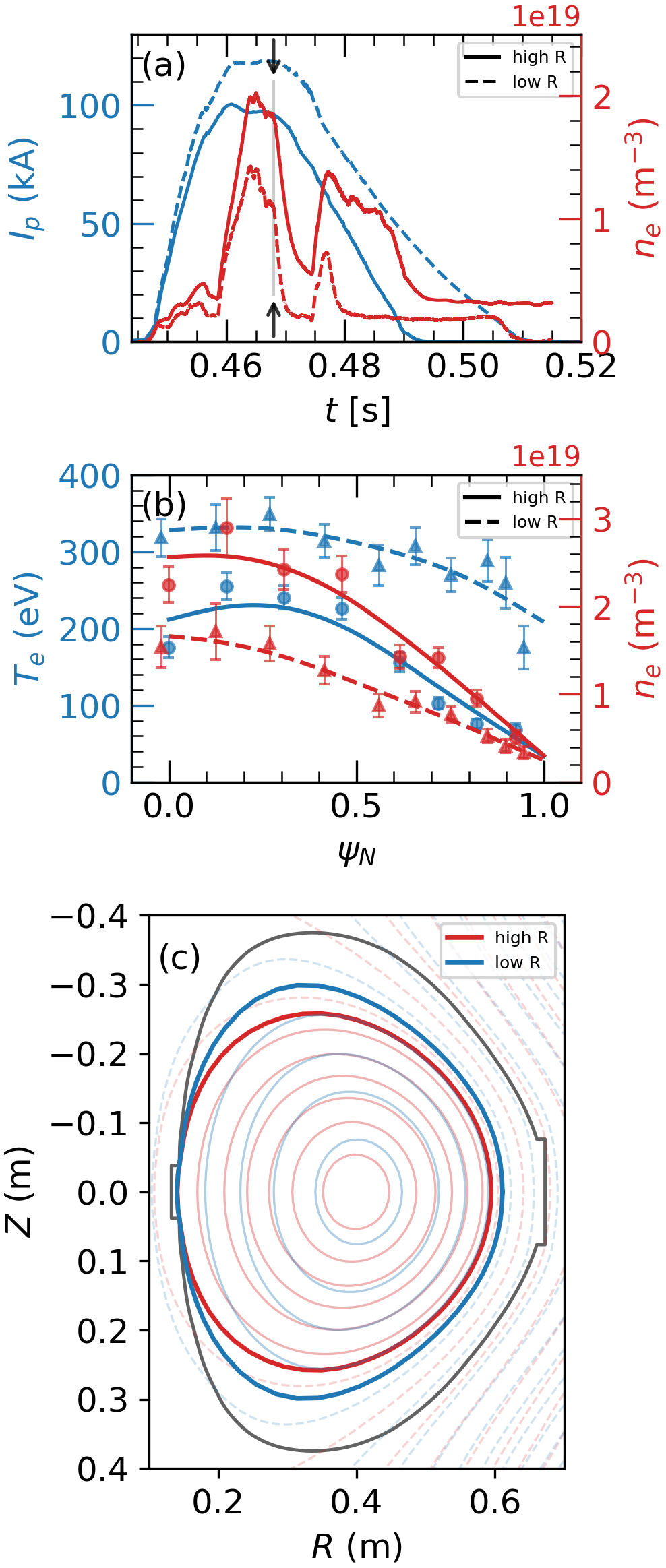}
  \caption[Shot comparison]{a) Plasma current and line averaged density for two representative shots, one with high recycling and one with low-recycling walls. The time point of investigation is indicated with a vertical line at t = 0.468 sec. b) Thomson scattering data for the two shots at t = 0.468 s. c) Kinetically (Thomson pressure profile) constrained equilibrium reconstruction of the two shots at the same time point.}
  \label{fig:shots}
\end{figure}

We compare two shots, 103795 and 103955 \cite{maan2024recycling}. These are the two median shots from ensembles of repeated identical shots from two run days, each with a different state of lithium on the walls. \added{Throughout this paper, the labels ``low-recycling case'' and ``high-recycling case'' are shorthand for the simulation parameter sets constrained by these experimentally characterized discharge ensembles; they do not imply that recycling is evolved by the model. Here, ``high recycling'' denotes the higher of the two \ltx recycling conditions rather than a conventional $R\simeq1$ high-recycling regime.} Shot 103795 will be referred to as the high-recycling case and is from a partially passivated lithium run day, where the previous lithium evaporation is 5 weeks old. It is understood from previous surface science analysis of the \ltx walls \cite{maan_ppcf_2021, maan_ieee_2020} that at least the top several nanometers of lithium is oxidized with a combination of lithium oxide and lithium hydroxide. Oxidized lithium is expected to have a higher hydrogen reflectivity and consequently lead to higher recycling \cite{krstic-li2o}. \deleted{This shot has a documented recycling coefficient of about 0.8} \added{A global recycling coefficient of $R=0.83\text{--}0.87$ was inferred for the passivated-lithium ensemble containing this shot} \cite{maan2024recycling}. Shot 103955 is the low-recycling case and has a fresh, nearly pure lithium evaporation of $\sim$ 863 mg, with an average coating thickness of $>$ 100 nm and \deleted{a documented recycling coefficient of about 0.5} \added{belongs to the ensemble for which a global recycling coefficient of $R=0.61\text{--}0.67$ was inferred} \cite{maan2024recycling}. 

The plasma current and line averaged density waveforms are shown in figure~\ref{fig:shots}(a), while electron temperature and density profiles from the core Thomson scattering system \cite{kaur2026ieee} are plotted in figure~\ref{fig:shots}(b). The two shots have identical gas fueling, but strong pumping by the lithium walls leads to a much lower line averaged density in the low-recycling shot. The core temperature is also higher in the low-recycling shot, likely due to improved confinement and higher plasma current from reduced recycling and reduced radiative losses compared to oxidized lithium. The PSI-TRI \cite{psi-tri} equilibrium reconstructions of the two shots appear in figure~\ref{fig:shots}(c). The open field line geometry is similar between the two shots, but the low-recycling shot has a larger plasma minor radius and elongation and therefore a slightly larger connection length along the SOL field line. We see that the edge density is similar between the two shots, but the edge temperature is higher in the low-recycling shot, consistent with reduced cooling from charge exchange with neutrals and a lower secondary emission coefficient of nearly pure lithium compared to oxidized lithium \cite{Capece2016see}. For purposes of the \gke simulations in section 3, the midplane density of both cases was taken to be $2\times10^{18}$ m$^{-3}$, but the temperature was set to the last measured point within the reconstructed boundary, 30 eV for the high-recycling case and 175 eV for the low-recycling case. Using the neoclassical collisionality expression \added{(equation~\ref{eq:nuStar})}, the electron collisionality of the two cases is  3.4 and 0.12; and the ion collisionality is 13.7 and 0.49.

\added{For comparison, the conventional upstream SOL electron collisionality is $\nu_{e,\mathrm{SOL}}^*=L_\parallel/\lambda_e\simeq10^{-16}n_{e,u}L_\parallel/T_{e,u}^2$, with density in m$^{-3}$, length in m, and temperature in eV~\cite{Stangeby2000}. Using the representative limiter-to-limiter connection length of 11 m~\cite{maan2024recycling} and the midplane parameters above gives $\nu_{e,\mathrm{SOL}}^*\simeq0.072$ and $2.44$ for the low and high-recycling cases, respectively. An analogous ion measure, $\nu_{i,\mathrm{SOL}}^*=L_\parallel\nu_{ii}/v_{ti}=L_\parallel/\lambda_{ii}$, gives $0.33$ and $12.9$, using $v_{ti}=\sqrt{T_{i0}/m_i}$, the ion self-collision frequency in equation~\ref{eq:nu_ss}, and reference ion temperatures of 70 and 10 eV, respectively. If the connection length is instead taken from the upstream midplane to one limiter, $L_\parallel=5.5$ m, the electron values are $0.036$ and $1.22$, and the ion values are $0.164$ and $6.47$. These SOL measures supplement the neoclassical collisionalities quoted above.}

\begin{figure}[htbp]
  \centering
  \includegraphics[width=0.45\textwidth]{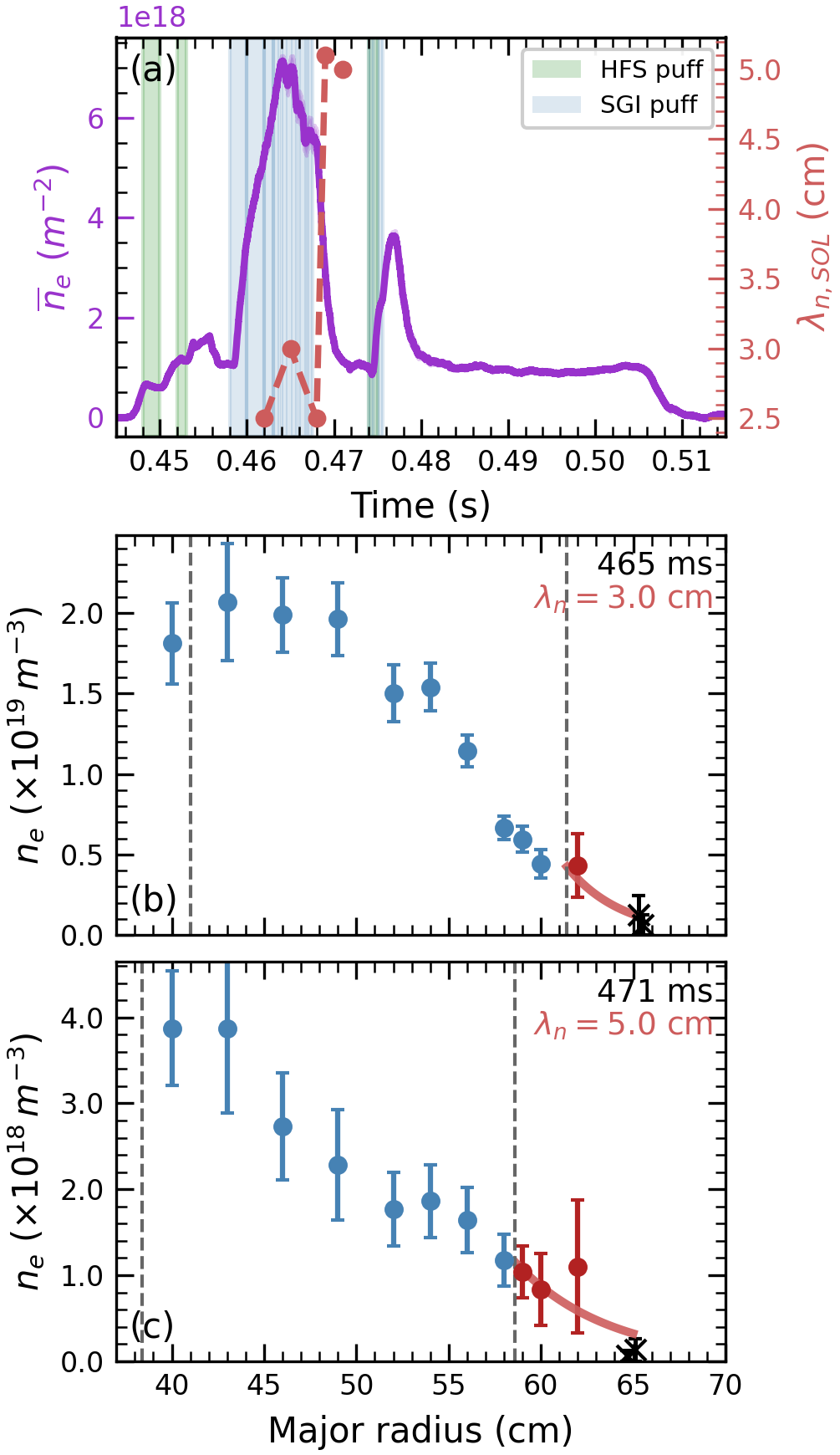}
  \caption{(a) Line-averaged density for the low-recycling shot, with the inferred low-field-side SOL density decay length $\lambda_{n,\mathrm{SOL}}$ overlaid on the right axis as a function of time. (b) Thomson scattering density measurement near peak line-averaged density, t = 465 msec, with the fitted profile and the LCFS anchor point from TRANSP and low field side probe points mapped to the midplane. (c) At t = 471 msec, after the SGI puff has terminated, with the same fit and anchor point. The inferred $\lambda_{n,\mathrm{SOL}}$ increases as the SOL warms up after the puff terminates.}
  \label{fig:lambdaNSOL_time}
\end{figure}

We also examined the evolution of the low-field-side density decay length during the low-recycling discharge. Figure~\ref{fig:lambdaNSOL_time} shows the inferred $\lambda_{n,\mathrm{SOL}}$, or radial density decay length in the SOL, as a function of time through the shot at five Thomson-scattering time points. The density decay length should be related to radial transport, and its evolution as a function of recycling/collisionality can provide indications about kinetic radial transport modifications. For the first three time points, the shot is being fueled by a supersonic gas injector (SGI). The SGI puff terminates about 200 microseconds before 468 msec, which is the time point corresponding to the temperature and density profiles in figure~\ref{fig:shots}. The estimate was constructed by stitching together several constraints at each time slice: the last-closed-flux-surface location from the kinetic equilibrium reconstruction, edge Thomson-scattering density points outside the LCFS, an LCFS anchor density from TRANSP, the two far-SOL Langmuir-probe density points marked by x symbols near the wall in figure~\ref{fig:lambdaNSOL_time}(b,c), projected back to the outboard midplane along open field lines. The fit is performed with a simple exponential function, $n_e(r)=n_{e0}\exp\left[-(r-r_{\mathrm{LCFS}})/\lambda_{n,\mathrm{SOL}}\right]$, where $n_{e0}$ is the density at the LCFS and $r-r_{\mathrm{LCFS}}$ is the radial distance outside the LCFS at the outboard midplane. The resulting $\lambda_{n,\mathrm{SOL}}$ is shown on the right axis of figure~\ref{fig:lambdaNSOL_time}(a) as a function of time, and the fits for two time points are given in figures~\ref{fig:lambdaNSOL_time}(b) and (c). We see that $\lambda_{n,\mathrm{SOL}}$ increases as the SGI puff terminates and the SOL warms up.

Since the fit combines measurements with different spatial coverage, uncertainties, and model dependence, the resulting $\lambda_{n,\mathrm{SOL}}$ inherits those systematic uncertainties. The probe densities are not used at their raw measurement location, but are mapped to the midplane using the reconstructed open field lines. Finally, the LCFS value used to anchor the fit is taken from TRANSP rather than measured directly by the edge Thomson system. These choices make the inferred decay length useful as a trend indicator, and only the temporal trend is interpreted here.

A similar analysis was not retained for the high-recycling case because the Thomson data and/or the fit quality was not sufficient for a comparable interpretation of the Thomson points past the LCFS. For the low-recycling case, the temporal trend should also be interpreted in the context of the gas puff. While the puff is active, the SOL is expected to remain relatively cold even with fresh lithium on the walls because the SGI floods the SOL with a cold neutral source. After the puff is terminated, the edge density falls and the colder population can be flushed out as the lithium walls continue to pump particles, allowing the SOL to warm back up \cite{maan2024recycling}. The increase in $\lambda_{n,\mathrm{SOL}}$ later in time likely reflects both the warming of a low-recycling SOL and a change in the balance between local fueling, wall pumping, and radial profile evolution. These experimental constraints motivate the field-line kinetic model described next.

\added{Kinetic studies of SOL parallel transport have examined non-Maxwellian distributions and sheath-connected losses across a wide range of collisionalities and recycling conditions~\cite{Procassini1990,Chodura1992,Manfredi2011,Takizuka2017}. Reduced electron conductive heat flux has also been obtained without magnetic mirroring~\cite{Batishchev1999,Zhao2017}. Experimentally constrained kinetic SOL modeling has been performed for the high-recycling COMPASS divertor~\cite{Tskhakaya2021}. Those conditions differ from the hot-edge regime expected when wall recycling and the associated neutral cooling are strongly reduced. This low-recycling regime, anticipated theoretically and demonstrated experimentally in LTX and LTX-$\beta$, provides the physical basis for the present study~\cite{Zakharov2004LiWall,Catto2006Isomak,Boyle2017PRL,maan2024recycling}. Building on preliminary kinetic modeling of these conditions, this study examines how mirror-trapped ion populations and the self-consistent kinetic-electron response jointly affect parallel density structure and effective SOL confinement. In addition to distribution functions and parallel profiles, we evaluate particle and energy residence times from the corresponding inventories and source rates. These diagnostics quantify how much plasma and thermal energy the SOL retains for a given fueling and heating rate, complementing the heat-flux and sheath diagnostics emphasized in earlier studies. Normalization to a sound-transit time distinguishes changes in effective confinement from the direct effect of temperature on parallel transit. Systematic collisionality scans, together with comparisons between kinetic and Boltzmann electrons and between varying and constant magnetic fields, connect changes in velocity-space structure to changes in these residence times. The combined analysis therefore examines how electron kinetics and ion mirror trapping modify particle and energy retention in experimentally accessible low-recycling lithium-wall conditions, with relevance to comparable regimes beyond LTX-$\beta$.}

\section{The \gke gyrokinetic model}

\begin{figure}[htbp]
  \centering
  \includegraphics[width=0.8\textwidth]{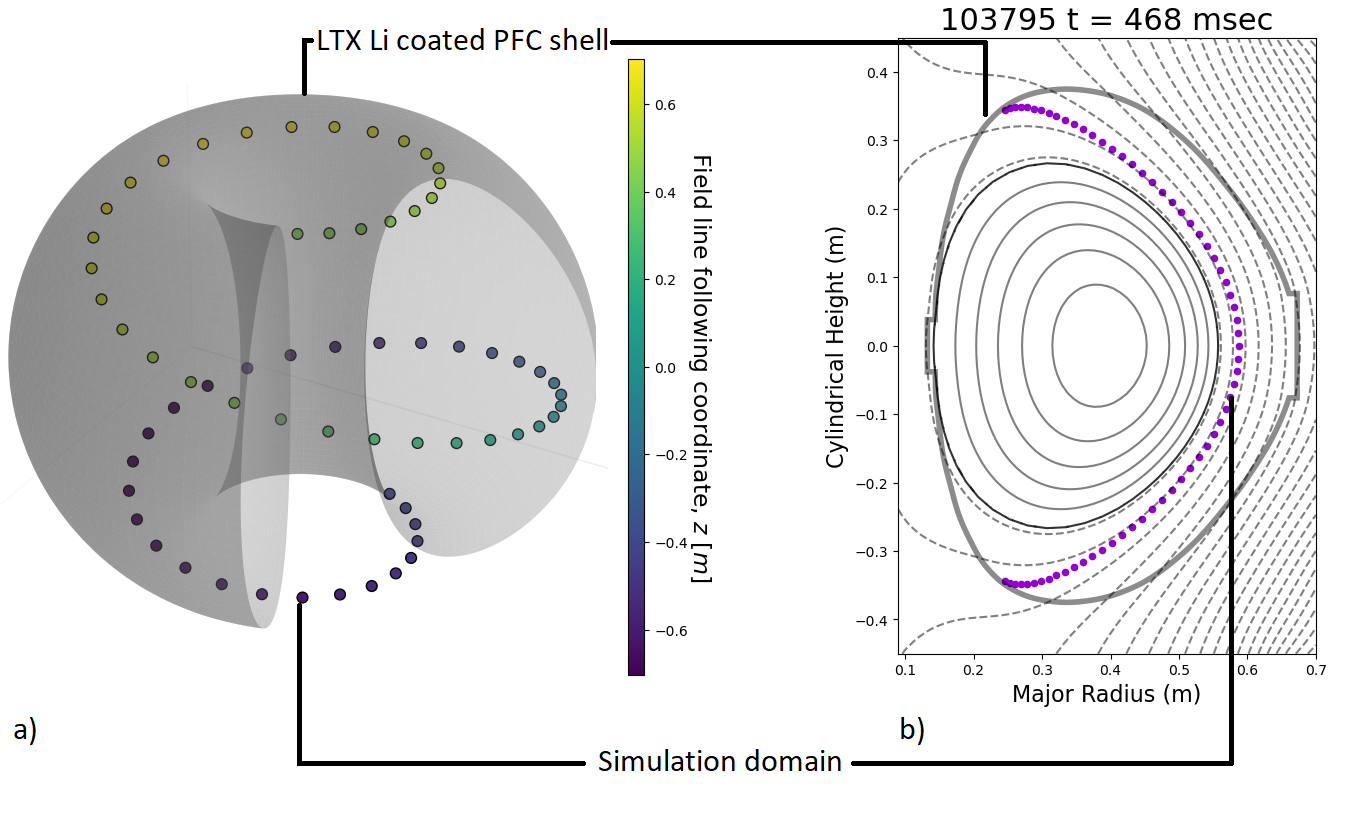}
  \caption[\gke field-line geometry.]{Field-line geometry used in the \gke simulations. The computational domain follows a single open field line that resembles the PSI-TRI reconstruction of the high-recycling discharge. The figure shows the location of the simulated field line relative to the LCFS for the high-recycling case, and the position of the LTX-$\beta$ shell. \textit{Reproduced from~\cite{maan2024recycling} under a Creative Commons Attribution (CC BY) license.}}
  \label{fig:gridPoints}
\end{figure}

\gke can solve a full-$f$ gyrokinetic model~\cite{Francisquez_2025} that obtains the time and phase-space evolution of the guiding center distribution function $f_s(\v{R},\vpar,\mu)$ of each species labeled $s$. Phase space here refers to the high-dimensional space spanned by the guiding center position ($\v{R}$), the particle velocity ($\vpar$) parallel to the equilibrium magnetic field $\v{B}=B\,\bhat$, and the magnetic moment ($\mu=m_s\vperp^2/(2 B)$) related to the perpendicular particle velocity $\vperp$. \gke can solve this equation in a variety of geometries and dimensionalities, and for this work we employ a field-aligned coordinate system~\cite{Beer1995} where the guiding-center position $\v{R}(\v{x})$ is mapped to the computational coordinates $\v{x}=(x,y,z)$:

\begin{equation} \label{eq:coords_comp}
    x = r-r_0,
    \qquad
    y = -\frac{r_0}{q_0}\left(\varphi+\alpha(r,\theta,\varphi=0)\right),
    \qquad
    z = \theta,
\end{equation}
where $\theta$ is the poloidal angle, $r_0$ is the minor radius at the center of the simulation domain, and $q_0$ is the safety factor at $r_0$ assuming an analytic profile $q(r)=q_{LCFS}/[1-s_{LCFS}(r-a_{OMP})/a_{OMP}]$ tailored to approximate and extend the $q$-profile inside the LCFS; here $a_{OMP}$ is the minor radius at the outboard mid-plane (OMP), and $q_{LCFS}$ and $s_{LCFS}$ are the safety factor and magnetic shear at the LCFS. The field-line label $\alpha=\alpha(r,\theta,\varphi)$ is calculated via
\begin{equation}
    \alpha = \varphi - I(\Psi)\int_0^\theta\mathrm{d}\theta'\,\frac{J_r(r,\theta')}{R(r,\theta')^2}, \qquad
    I(\Psi) = 2\pi q\left(\int_0^{2\pi}\mathrm{d}\theta\,\frac{J_r(r,\theta)}{R(r,\theta)^2}\right)^{-1}.
\end{equation}
The integrands contain the Jacobian $J_r = R\left[\left(\partial_r R\right)\partial_\theta Z-\left(\partial_\theta R\right)\partial_r Z\right]$ of the $(r,\theta,\varphi)$ coordinates, which can be computed for a Miller model of the flux surfaces~\cite{Miller1998}
\begin{eqnarray}
    R(r,\theta) &=& \Raxis + r\cos\left(\theta + \arcsin(\delta)\sin(\theta)\right), \\
    Z(r,\theta) &=& \kappa r\sin(\theta)
\end{eqnarray}
where $\delta$ and $\kappa$ are the triangularity and elongation, respectively. We adjusted this analytic equilibrium, using $\Raxis=0.4$, $\kappa=1.57$, $\delta=0.6$, to give an open-flux surface that resembles that inferred by a kinetically constrained PSI-TRI \cite{psi-tri} equilibrium (see figure~\ref{fig:gridPoints}(b)). Also informed by equilibrium reconstruction is our choice of $q_{\tLCFS}=3.6955$, $s_{\tLCFS}=2.2798$ and $\Baxis=0.2244$ T.

Since we are interested in characterizing parallel dynamics of the \ltx~SOL, we use \gke in the electrostatic limit in one position-space dimension along a single SOL field line crossing the OMP at $r_0=R_0-\Raxis$ or $R_0=R_{\tLCFS,\tOMP}+0.025\,\mathrm{m}$ (i.e. 2.5 cm outside the LCFS whose major radius at the OMP is $R_{\tLCFS,\tOMP}=0.5948$ m). \gke is thus used here to evolve (in time) the guiding center distribution, $f_s(z,\vpar,\mu)$, of a species $s$ with mass $m_s$ and charge $q_s$ as a function of the along-the-field-line coordinate $z$ (as well as $\vpar$ and $\mu$). Specifically, we solve
\begin{equation} \label{eq:gkeq1d}
    \pd{\phaseJac f_s}{t} + \frac{1}{\geoJac}\pd{}{z}\left(\geoJac\dot{z}\,\phaseJac f_s\right) + \pd{}{\vpar}\left(\dot{\vpar}\,\phaseJac f_s\right) = \phaseJac \added{\mathcal{C}_s[f_s]} + \phaseJac S_s.
\end{equation}
The time derivatives of $z$ and $\vpar$ are
\begin{equation}
    \dot{z} = \frac{\cmag}{\geoJac\Bstarpar}\vpar,
    \qquad
    \dot{\vpar} = -\frac{\cmag}{\geoJac\Bstarpar}\frac{1}{m_s}\left(\mu\pd{B}{z}+q_s\pd{\phi}{z}\right),
\end{equation}
where $\phaseJac=\Bstarpar=\bhat\cdot\vBstar\approx B$ is the Jacobian of the guiding-center transformation, and $\phi$ is the electrostatic potential. 
The function $\geoJac=\sqrt{\mathrm{det}\left(g_{ij}\right)}$ corresponds to the Jacobian of the $\v{R}(\v{x})$ transformation with the covariant metric tensor $g_{ij}$, such that $\cmag=\geoJac\bmag/\sqrt{g_{zz}}$. The magnetic field magnitude $\bmag=\left(B_{\varphi}^2+B_{\theta}^2\right)^{1/2}$ is composed of the toroidal $B_{\varphi}(R)=\Baxis\Raxis/R$ and poloidal $B_\theta=\abs{\grad{\varphi}\times\grad{\Psi}}=\abs{\grad{r}}\Psi'/R$ components, where $\Psi'=B_\varphi(R_0)\Raxis/I(\Psi)$ and $\abs{\grad{r}}^2=\left(R/J_r\right)^2\left[\left(\partial_\theta R\right)^2+\left(\partial_\theta Z\right)^2\right]$.

\deleted{On the right-hand side of equation~\ref{eq:gkeq1d}, the $\mathcal{C}[f_s]$ stands for the collision operator, which in this model is a Dougherty operator with the form}
\added{On the right-hand side of equation~\ref{eq:gkeq1d}, $\mathcal{C}_s[f_s]$ is the multispecies gyroaveraged Dougherty collision operator,}
\begin{equation} \label{eq:lbo}
    \added{\mathcal{C}_s[f_s] = \sum_r \nu_{sr}\left\{
    \pd{}{\vpar}\left[\left(\vpar-u_{\parallel,sr}\right)f_s+v_{t,sr}^2\pd{f_s}{\vpar}\right]
    +\pd{}{\mu}\left[2\mu f_s+2\frac{m_s v_{t,sr}^2}{B}\mu\pd{f_s}{\mu}\right]\right\}.}
\end{equation}
\deleted{where $\upars$ and $v_{ts}$ are the parallel drift and thermal speeds, respectively, calculated with a prescription that conserves momentum and energy to machine precision~\cite{Francisquez2020}, and the collision frequency is calculated with~\cite{Huba2019nrl}}
\added{Here $r$ denotes the collision partner, and the sum includes both self- and unlike-species collisions. For $r=s$, $u_{\parallel,ss}=u_{\parallel s}$ and $v_{t,ss}^2=v_{ts}^2$. For $r\ne s$, $u_{\parallel,sr}$ and $v_{t,sr}^2$ are the conservative cross-species primitive moments. These moments are calculated from the evolving distributions using prescriptions that conserve momentum and energy to machine precision~\cite{Francisquez2020,Francisquez2022}. Electron--ion and ion--electron collisions were enabled in all simulations with kinetic electrons. The self-collision frequencies ~\cite{Huba2019nrl} are calculated using}
\begin{equation} \label{eq:nu_ss}
\begin{aligned}
    \log\Lambda_s &= 6.6 - 0.5 \log\left(n_{s0}/10^{20}\right) + 1.5 \log\left(T_{s0}/e\right), \\
    \nu_{ss} &= \nu_{\mathrm{frac}} \frac{e^4 n_{s0}\log\Lambda_{s}}{3 \left(2\pi\right)^{3/2} \epsilon_0^2 \sqrt{d_sm_s}~T_{s0}^{3/2}}, \\
\end{aligned}
\end{equation}
where \added{$n_{s0}$ and $T_{s0}$ are the fixed reference number density and temperature (in energy units) of species $s$,} $\nu_{\mathrm{frac}}=1$, $d_s=1$ for electrons and $d_s=2$ for ions. \added{For the \textsc{Gkeyll} version used here, the cross-species frequencies are $\nu_{ei}=\sqrt{2}\nu_{ee}$ and $\nu_{ie}=(m_e/m_i)\nu_{ei}$. Note that the Dougherty operator in equation~\ref{eq:lbo} lacks the 1/$v^3$ dependence in the collision frequency characteristic of the full Fokker-Planck operator. This simplification removes physics, like the thermal force, that can be important in SOL transport~\cite{Homma2020}, and reduces the accuracy with which we model heat-carrying particles that are important in setting divertor heat loads, especially in collisionless reactor conditions~\cite{Chankin2015}.} 

\begin{figure}[htbp]
  \centering
  \includegraphics[width=\textwidth]{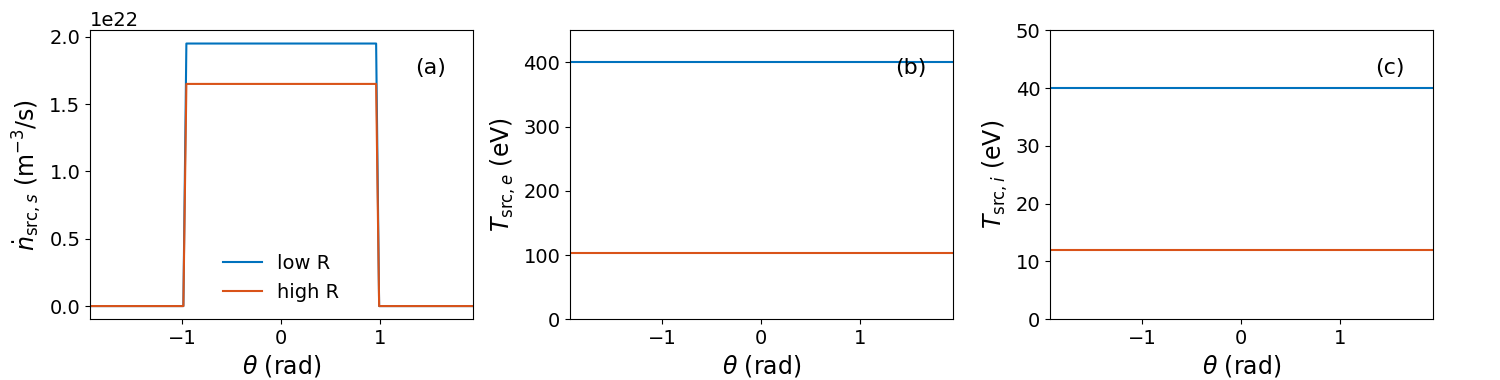}
  \caption{(a) Density source rate, (b) electron source temperature, and (c) ion source temperature for low (blue) and high (orange) \recR~simulations.}
  \label{fig:src_mom}
\end{figure}

\begin{figure}[htbp]
  \centering
  \includegraphics[width=\textwidth]{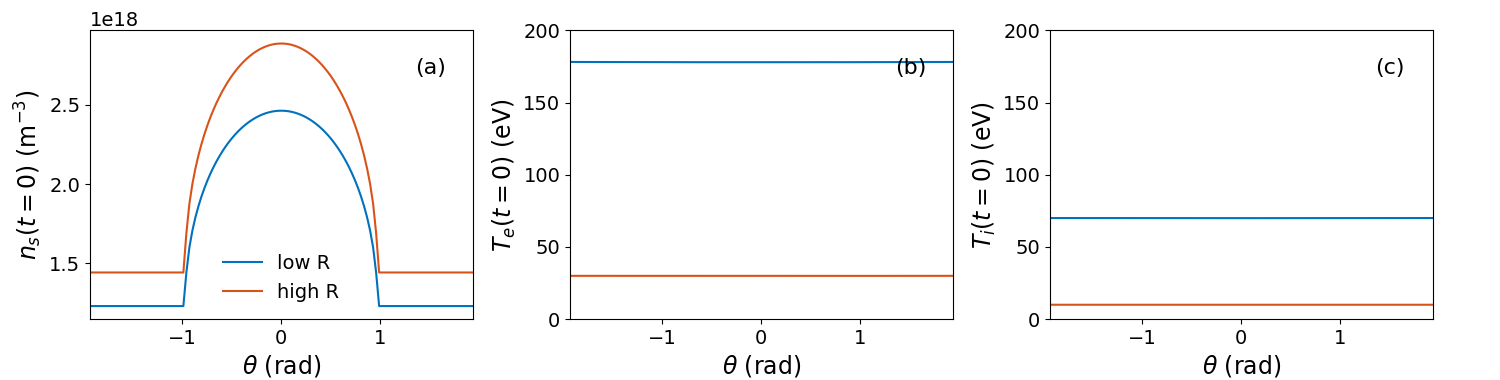}
  \caption{(a) Density, (b) electron temperature, and (c) ion temperature of the non-drifting Maxwellian initial condition (IC) for low (blue) and high (orange) \recR~simulations.}
  \label{fig:ic_mom}
\end{figure}

Equation~\ref{eq:gkeq1d} also contains a source term, $S_s$, that stands as a proxy for radial transport from the core into the SOL and for particle sources due to recycling. Such a source has a Maxwellian shape in velocity space,
\begin{equation}
    \mathcal{S}_s = \frac{\dot{n}_{\mathrm{src}}}{\left(2\pi T_{\mathrm{src},s}/m_s\right)^{3/2}}\exp\left[-\frac{\vpar^2+2\mu B/m_s}{2T_{\mathrm{src},s}/m_s}\right],
\end{equation}
with the density source rate ($\dot{n}_{\mathrm{src}}$) and source temperature ($T_{\mathrm{src},s}$) as in figure~\ref{fig:src_mom}. Since these rates and temperatures influence the final density and temperature, we pick these approximately so as to produce densities and temperatures for each of the low and high \recR~cases informed by the measurements presented in figures~\ref{fig:shots}-\ref{fig:lambdaNSOL_time}. These final profiles are a result of evolving the non-drifting Maxwellian initial conditions (ICs), with the moments in figure~\ref{fig:ic_mom}, according to equation~\ref{eq:gkeq1d}, using zero-flux boundary conditions (BCs) in velocity space and conducting sheath BCs at the ends of the field line~\cite{Shi2017thesis}, until little to no change is observed. 

In order to close the system of equations we need a field equation that determines the evolution of the electrostatic potential. In this work we employed two models used in previous 1D SOL modeling, an approximation to the gyrokinetic Poisson equation, and an Boltzmann electron model.

\subsection{One-dimensional approximation to the Gyrokinetic Poisson equation} \label{sec:fieldModel1d}

In the long wavelength limit the gyrokinetic Poisson equation is~\cite{Shi2017}
\begin{equation} \label{eq:poisson3d}
    -\div{\sum_s\frac{m_sn_{0s}}{B_0^2}\gradperp{\phi}} = \sum_sq_s\frac{2\pi}{m_s}\int \dvpar\,\dmu~\phaseJac f_s,
\end{equation}
where we have made use of a Boussinesq approximation to use the reference density $n_{0s}$ and magnetic field amplitude $B_0$ on the left side of this equation, and $\gradperp{}$ is a gradient perpendicular to the magnetic field. \added{The Boussinesq approximation has of course been identified as a source of error in 2D blobby transport~\cite{Angus2014,Held2023}, although that error seems to become small in 3D~\cite{Angus2014,Ross2018}.} In this work, however, we pursue one-dimensional simulations along the field line and the perpendicular direction is  assumed ignorable. Yet, in equation~\ref{eq:poisson3d} the variation of the potential $\phi$ along the field line is given by the distribution of charge on a given perpendicular plane. Although there are other candidates for field equations in 1D models, due to its simplicity, we opt for an approximation to the above Poisson equation in which we assume a Fourier transformation in the perpendicular plane and replace the $\div{\gradperp{}}$ operator with an assumed perpendicular wavenumber $\kperp$~\cite{Shi2015,Geyko2024}. Our one-dimensional field equation thus becomes
\begin{equation} \label{eq:poisson1d}
    \kperp^2\sum_s\frac{m_sn_{0s}}{B_0^2}\phi = \sum_sq_s\frac{2\pi}{m_s}\int \dvpar\,\dmu~\phaseJac f_s.
\end{equation}
Formally in a 1D model $\kperp$ should go to zero, but in order to use a limit of equation~\ref{eq:poisson3d} to obtain the electric field along the field line we need to keep an approximation to the perpendicular operators on the left side. This model requires us to specify a $\kperp$ (or, more conveniently, $\kperp\rho_{s0}$, for the reference ion sound gyroradius $\rho_{s0}=\sqrt{m_iT_{e0}}/(q_iB_0)$ in terms of the reference electron temperature $T_{e0}$). We will evaluate the sensitivity of our results to this free parameter and ensure that we have a sufficiently low $\kperp\rho_s$ to approximate the asymptotic limit.

\subsection{Boltzmann electrons with ambipolar sheath fluxes} \label{sec:boltz}

As a point of reference alternative to the gyrokinetic Poisson model in section~\ref{sec:fieldModel1d}, we also pursued simulations with a framework that assumes Boltzmann electrons and ambipolar sheath fluxes~\cite{Shi2015,Francisquez2023}. The equality of electron ($\Gamma_{es}$) and ion ($\Gamma_{is}$) particle fluxes at the sheath entrance ($z=z_s$) means
\begin{equation}
\Gamma_{es}=\frac{2\pi}{m_e}\int\dvpar\dmu\,\vpar\,\phaseJac\left\{\frac{n_e}{\left(2\pi \added{v_{te0}^2}\right)^{3/2}}\exp\left[-\frac{\vpar^2+2\mu B/m_e}{2v_{te0}^2}-\frac{e\phi_s}{T_{e0}}\right]\right\}\Bigg|_{z=z_s} = \Gamma_{is},
\end{equation}
where $v_{te0}=\sqrt{T_{e0}/m_e}$ is the thermal speed of isothermal electrons. Carrying out this integral and assuming quasineutrality leads to a prescription of the sheath potential:
\begin{equation} \label{eq:sheathpot}
    \phi_s = \frac{T_{e0}}{e}\ln\left(\frac{n_{is}v_{te0}}{\sqrt{2\pi}\,\Gamma_{is}}\right);
\end{equation}
here we denoted the ion density at the sheath entrance as $n_{is}$. The sheath potential is employed in the Boltzmann electron response
\begin{equation} \label{eq:neBoltz}
    n_e(z) = n_{es}\exp\left[\frac{e\left(\phi(z)-\phi_s\right)}{T_{e0}}\right],
\end{equation}
in terms of the electron density at the sheath entrance $n_{es}$. From equation~\ref{eq:neBoltz} we can obtain the variation of the potential along the field line:
\begin{equation} \label{eq:adiabaticphi}
     \phi(z) = \phi_s+\frac{T_{e0}}{e}\ln\left(\frac{n_i(z)}{n_{is}}\right),
\end{equation}
where we once again employed quasineutrality to replace $n_{es}=n_{is}=n_i(z=z_s)$. In a time-dependent simulation we can thus solve for the ion dynamics, compute the ion density ($n_i$) and particle flux at the sheath entrance ($\Gamma_{is}$) and use equations~\ref{eq:sheathpot} and~\ref{eq:adiabaticphi} to obtain the potential along the field line. Since this model does not evolve kinetic electrons, it provides an alternative approximation at significantly lower computational cost (around 42X cheaper for hydrogen ions).

\section{Simulation results} \label{sec:results}

The simulations presented here can be reproduced with the open-source code Gkeyll and the provided input files, see~\ref{sec:getGkeyll}. We begin by running reference simulations based on the experimental data described in section 2 and comparing the high-recycling and low-recycling cases. To assess the importance of kinetic-electron effects, we run the simulations with both kinetic-electron and Boltzmann-electron models. These runs establish the baseline density, flow, and temperature structure along the modeled SOL field line. The remainder of the section uses these reference cases (unless otherwise noted) to motivate the collisionality and magnetic-field scans discussed later. We then analyze the kinetic effects by examining the ion distribution function, the non-Maxwellian component of the \deleted{electron} \added{ion} distribution function, and the trapped particle fraction. Finally, we present the results of collisionality scans for both cases, showing how the density profiles and SOL particle and energy confinement times change as a function of collisionality.

The equilibrium geometry, source profile, and numerical resolution are prescribed as run inputs. \gke is initialized with edge density and temperature values of approximately $n_e = 2\times10^{18}\,\mathrm{m}^{-3}$ for both the high-recycling and low-recycling cases, $T_e = 175$ eV for the low-recycling case, and $T_e = 30$ eV for the high-recycling case. An initial ion temperature of $T_i = 0.4\, T_e$ eV is used. \gke then evolves the initially Maxwellian ion and electron distribution functions self-consistently with the electrostatic potential along a single open field line. The resulting dynamics therefore reflect parallel kinetic transport in a fixed magnetic geometry, together with the effects of sources, sheath losses, and collisions. The sources are localized near the midplane and are designed to match the experimental midplane density and temperature. The sheath boundary conditions at the limiter end of the field line allow for kinetic losses of both ions and electrons, and the limiter is assumed to be perfectly absorbing. The numerical resolution is chosen based on the convergence studies summarized in ~\ref{sec:convergence}. 

\subsection{Reference simulations} \label{sec:baseSims}

The reference simulations were obtained by solving equation~\ref{eq:gkeq1d} with the Gkeyll code~\cite{gkeyll}, using a discontinuous Galerkin discretization with $(\Nz,\Nvpar,\Nmu)=(64,16,12)$ cells along $(z,\vpar,\mu)$, respectively, and a piecewise linear basis along $z$ and $\mu$ but a piecewise quadratic basis along $\vpar$~\cite{Francisquez_2025}. The length of the domain along $z$ is simply the connection length for the selected field line (see figure~\ref{fig:gridPoints}), while in velocity space the grid extents are $\vpar\in\left[-4v_{ts0},4v_{ts0}\right]$ and
$\mu\in\left[0,m_s\left(6v_{ts0}\right)^2/(2B_0)\right]$, where $v_{ts0}=\sqrt{T_{s0}/m_s}$ is the reference thermal speed of species $s$. Note that we use a nonuniform velocity discretization, using a quadratic grid in $\mu$ and a linear-quadratic grid in $\vpar$~\cite{Francisquez_2025}. A resolution convergence study is shown in~\ref{sec:convergence}.

\begin{figure}[htbp]
  \centering
  \includegraphics[width=0.65\textwidth]{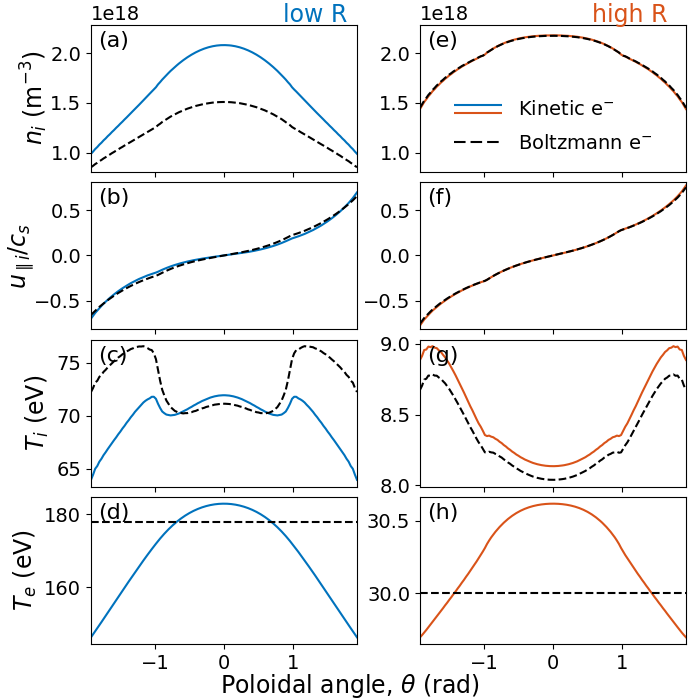}
  \caption{(a) Ion number density, (b) parallel speed (normalized to $c_s=\sqrt{(\Tpare+3\Tpari)/m_i}$), (c) ion temperature, and (d) electron temperature for the low-recycling case. Panels (e)--(h) show the same quantities for the high-recycling case. The kinetic-electron results are shown in solid lines, while the Boltzmann-electron results are shown in dashed lines.}
  \label{fig:nTprofiles}
\end{figure}

The results show that the low-recycling case sustains a hotter temperature than the high-recycling case. The midplane density evolves to similar values, with the high-recycling case showing a slightly higher value than the initial choice of $n_e = 2\times10^{18}\,\mathrm{m}^{-3}$, as shown in figure~\ref{fig:nFluxTT}. However, the hotter low-recycling case develops a steeper density gradient along the modeled field line. The ratio of midplane to limiter density is also modified for the low-recycling case compared to the high-recycling case, as shown in figure~\ref{fig:nTprofiles}.

\added{The conducting-sheath boundary condition reflects electrons below the sheath-energy cutoff and absorbs outgoing ions, but does not resolve the Debye sheath or explicitly impose a Bohm flow speed~\cite{Shi2017thesis}. The subsonic boundary flows have not been verified against the kinetic Bohm criterion. Numerical sensitivity of the sheath-entrance flow is discussed further in \ref{sec:constBsims}.}

The simulations were allowed to evolve toward a quasi-steady state. Time traces of the moments at the midplane, an intermediate location, and the sheath entrance are shown in ~\ref{sec:convergence}. Numerical convergence checks for velocity-space resolution, velocity-space extent, and the prescribed $\kperp\rho_s$ used in the 1D field model are also shown there. The electrostatic potential is somewhat more sensitive to the assumed $\kperp\rho_s$ than the plasma moments, so we show that scan explicitly here. Reference simulations used a value of $\kperp\rho_s=0.15$, at which point the electrostatic potential at the midplane, normalized to the electron temperature, is within 1\% of the asymptotic value (at $\kperp\rho_s=0.025$).

\begin{figure}[htbp]
  \centering
  \includegraphics[width=0.85\textwidth]{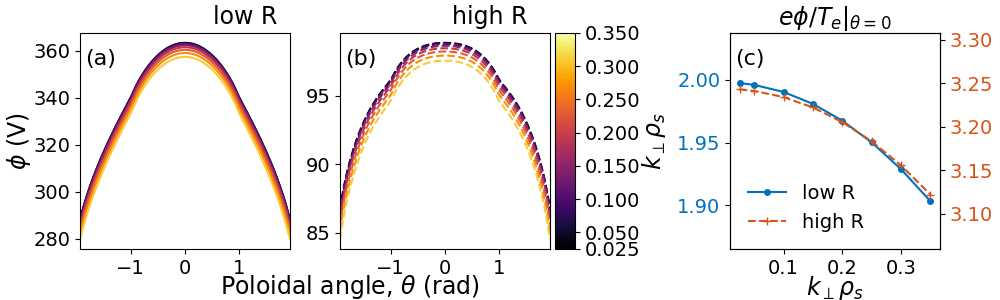}
  \caption{Electrostatic potential profiles for simulations of the low-recycling case (a) and high-recycling case (b) as a function of the parameter $\kperp\rho_s$, and the central potential normalized to the central electron temperature for each value of $\kperp\rho_s$ (c).}
  \label{fig:resConvergence_kperp_phiDTe}
\end{figure}

Figure~\ref{fig:resConvergence_kperp_phiDTe} shows that the electrostatic potential differs substantially between the two recycling states. The low-recycling case develops a much larger potential and a much larger variation of that potential along the field line than the high-recycling case. In a mirror geometry, particles are trapped when conservation of total energy and magnetic moment causes their parallel velocity to vanish before they can reach the high-field region, producing a trapped-passing boundary in velocity space \cite{Pastukhov1974,Ochs2023mirror}. In the LTX-$\beta$ SOL, once the collisionality falls well below unity, the magnetic-field variation along the field line provides that mirror-trapping geometry. In the high-recycling case, this distinction is less important because the collisionality is large enough that trapped ion orbits are expected to be disrupted by pitch angle scattering before mirror trapping can strongly structure the distribution function. In that regime, the dynamics remain closer to the conventional high-recycling SOL picture, where collisional transport, parallel flow, and sheath-connected losses dominate.

In the low-recycling case, however, the larger potential implies a stronger parallel electric field, directed from the midplane toward the limiter on each side of the domain, since the potential peaks near the midplane and falls toward the sheath entrance. That field exerts a force on the electrons back toward the midplane and therefore acts to reduce direct electron end loss, while ions are more readily accelerated toward the limiter. This is the ambipolar response required to help restore quasineutrality in a system where electrons would otherwise escape too efficiently along the open field line. At the same time, the low-recycling case is also hotter, so although the potential is much larger in volts, the normalized quantity $e\phi/T_e$ is not larger than in the high-recycling case. Taken together, these results suggest that the low-recycling LTX-$\beta$ SOL is not well described by standard collisional SOL fluid descriptions alone. Instead, it enters a regime in which weak collisionality, magnetic mirror trapping, and the ambipolar potential combine to produce a more complex SOL structure with stronger parallel gradients and non-Maxwellian features in the distribution functions. The rest of this section explores these kinetic effects in more detail and shows how they evolve as the collisionality is varied.

\added{The boundary potentials represent the unresolved sheath drops relative to the wall. Using the local sheath-entrance electron temperatures of approximately 145 and 30 eV gives $e\phi_s/T_{e,s}\simeq1.9$ and $2.8$ for the low and high-recycling cases, respectively. We interpret the smaller normalized sheath drop in the low-recycling case in terms of the combined action of the ambipolar potential along the field line and the sheath in regulating electron losses. An extended Pastukhov-type ambipolar potential was previously anticipated for the weakly collisional, mirror-trapped LTX SOL~\cite{Majeski2017}. Depletion of the higher-energy electron tail can further reduce the sheath potential required to balance electron and ion losses.}

Conventional SOL fluid models \cite{Stangeby2000} provide a useful reference point for interpreting the present simulations. In the sheath-limited regime, parallel losses to the material boundary dominate and the plasma can often be approximated as nearly isothermal along the field line, with the sheath setting the loss condition at the wall. In the conduction-limited or high-recycling regime, stronger parallel temperature gradients develop and thermal conduction plays a larger role in setting the upstream-to-target connection. These fluid descriptions capture the basic balance between parallel transport, cross-field source terms, and sheath losses, but they rely on closure assumptions that become less reliable as the SOL becomes weakly collisional and non-Maxwellian. The 1x2v \gke model used here retains that same open-field-line geometry while allowing the particle distributions and electrostatic response to evolve kinetically, making it better suited for assessing departures from standard fluid behavior in the low-recycling LTX-$\beta$ SOL.

A Boltzmann-electron closure was explored to isolate the role of kinetic electron dynamics in these simulations. In this limit, the electrons respond adiabatically to the electrostatic potential and do not develop the full non-Maxwellian structure allowed in the kinetic model. Comparing the Boltzmann and kinetic-electron cases therefore helps identify which features of the density, flow, and temperature profiles arise from ion kinetics alone and which depend on kinetic electron effects. Figure~\ref{fig:nTprofiles} shows that, for the low-recycling case, the density profile is underpredicted when adiabatic Boltzmann electrons are used. The trend suggests that an adiabatic electron response is not sufficient to capture the parallel force balance in the low-recycling SOL. Instead, the kinetic-electron result is consistent with a Pastukhov-type ambipolar potential \cite{Pastukhov1974} developing along the field line, together with the associated parallel electric field needed to reduce electron end loss and maintain quasineutrality. In the high-recycling case, where the kinetic and Boltzmann-electron profiles nearly overlap, those additional kinetic-electron effects appear to be much less important. 

\subsection{Changes with collisionality} \label{sec:effects}

\begin{figure}[htbp]
  \centering
  \includegraphics[width=0.5\textwidth]{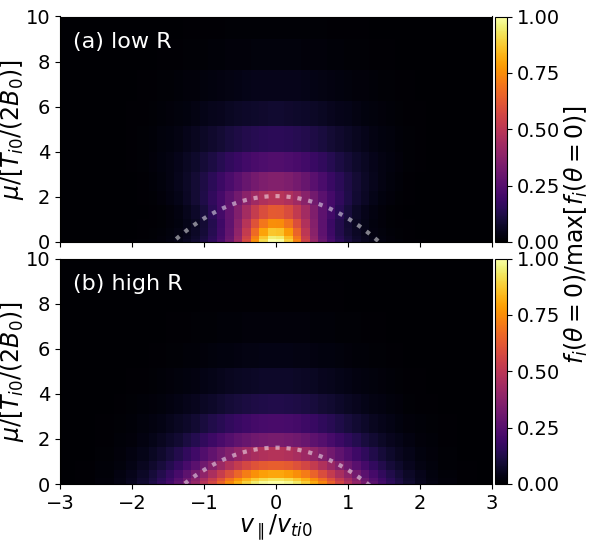}
  \caption{Ion distribution function at \added{the outboard midplane,} $\theta=0$, for simulations of the low-recycling case (a) and high-recycling case (b). \added{This is a local velocity-space slice, not an average along the field line.} White dotted lines indicate the $m_iv^2/2=T_i(\theta=0)$ contour in order to compare to an equivalent Maxwellian distribution of the same temperature.}
  \label{fig:distFion}
\end{figure}

\begin{figure}[htbp]
  \centering
  \includegraphics[width=0.5\textwidth]{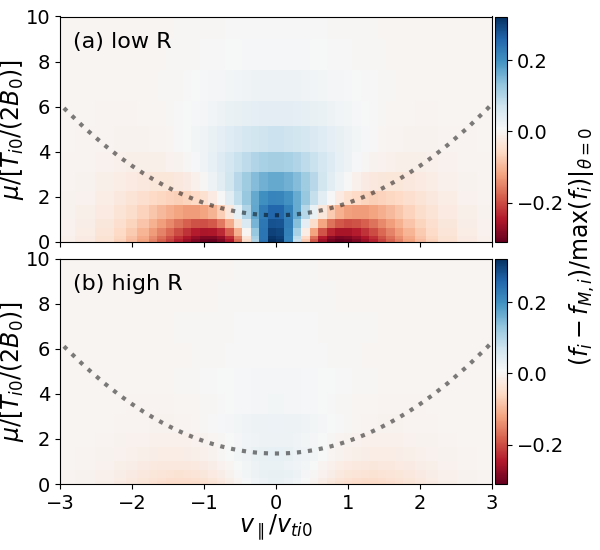}
  \caption{Non-Maxwellian component of the \deleted{electron} \added{ion} distribution function for the low-recycling case (a) and high-recycling case (b), at the \added{outboard} midplane ($\theta=0$). \added{The dotted line marks the trapped--passing boundary defined by equation~\ref{eq:lossCone}.}}
  \label{fig:nonMaxwellian}
\end{figure}

The reference simulations show that in the low-collisionality limit, the SOL departs most strongly from standard fluid-like behavior. To examine that regime more closely, we next vary the collisionality numerically and track how the phase-space structure responds. We begin with the ion distribution function and the non-Maxwellian component of the \deleted{electron} \added{ion} distribution at the midplane, shown in figures~\ref{fig:distFion} and~\ref{fig:nonMaxwellian}, which show stronger non-Maxwellian structure in the low-recycling case. We then quantify the trapped fraction and connect it to the low-collisionality behavior of the SOL poloidal density profiles, poloidal density peaking, and the SOL particle and energy confinement times. The collisionality scan is used to identify how mirror trapping and kinetic effects emerge and become dynamically important in the weakly collisional low-recycling SOL. 

\added{The distributions in figures~\ref{fig:distFion} and~\ref{fig:nonMaxwellian} are sampled locally at the outboard midplane. Away from the midplane, parallel electric and mirror forces shift and reshape the distribution in $\vpar$, so its departure from a Maxwellian varies along the field line. Spatial variation between mirror-trapped upstream distributions and skewed downstream distributions has also been demonstrated in other Gkeyll SOL simulations~\cite{Shukla2025}.}

We define the collisionality used in the scan as the standard neoclassical collisionality~\cite{HelanderSigmar2002}.
\begin{equation} \label{eq:nuStar}
\nu_i^* = \frac{\nu_{ii}/\epsilon}{\omega_{bi}}
= \frac{\nu_{ii} q_0 R_0}{\epsilon^{3/2} v_{ti}},
\end{equation}
where $\nu_{ii}$ is the ion-ion collision frequency, $v_{ti}$ is the ion thermal speed, and $\omega_{bi}=\sqrt{\epsilon}\,v_{ti}/(q_0R_0)$ is the ion bounce frequency. Here $\epsilon=r_0/R_0$ is the inverse aspect ratio evaluated at the outboard midplane location of the simulated field line, $R_0$ is the major radius of that field line at the outboard midplane, $r_0=R_0-R_{\rm axis}$ is the corresponding minor radius, and $q_0$ is the safety factor at $r_0$, from the kinetic reconstruction of the shot at that time point. In the collisionality scan, the collision frequencies are varied by a multiplicative factor $\nu_{\mathrm{frac}}$ \added{defined in equation~\ref{eq:nu_ss}}, so that the plotted values of $\nu_i^*$ correspond to the value $\nu_{\mathrm{frac}}\nu_i^*$.

\begin{figure}[htbp]
  \centering
  \includegraphics[width=0.5\textwidth]{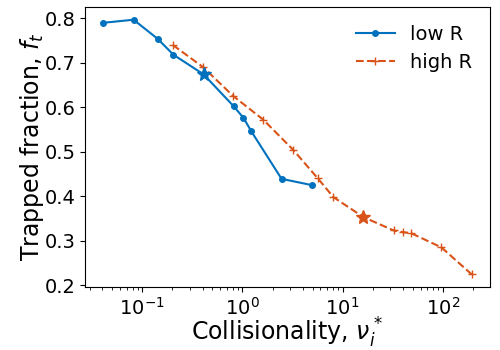}
  \caption{Trapped fraction as a function of ion collisionality for simulations of the low-recycling case (solid blue) and high-recycling case (dashed orange). \added{The base cases, with unscaled collision frequencies, are indicated by stars.}}
  \label{fig:fTrap_nuStar}
\end{figure}

\added{
\begin{equation}
\mu_{\mathrm{lc}}(v_{\parallel})  = \frac{1}{B_{min}}\,\frac{\frac{1}{2} m_s v_{\parallel}^2+q_s\left(\phi(\theta=0) - \phi(\theta_{\max})\right)}{R_m - 1}=\frac{m_s v_{\parallel}^2+2q_s\left(\phi(\theta=0) - \phi(\theta_{\max})\right)}{2 B_{min} (R_m - 1)}
\label{eq:lossCone}
\end{equation}}
The trapped fraction is calculated from the midplane ion distribution function by using equation~\ref{eq:lossCone} to separate trapped and passing regions in $(v_{\parallel},\mu)$ space. For each simulation in the collisionality scan, the magnetic-field magnitude along the field line is used to define $B_{min}=\min B$, $B_{max}=\max B$, and the mirror ratio $R_m=B_{max}/B_{min}$. The distribution function is then sampled at the midplane, and all phase-space cells with $\mu>\mu_{\mathrm{lc}}(v_{\parallel})$ are counted as trapped. The trapped fraction plotted in figure~\ref{fig:fTrap_nuStar} is the ratio of the distribution-function weight in this trapped region to the total distribution-function weight at the midplane. This diagnostic therefore measures how much of the local ion distribution occupies mirror-trapped phase space as the ion neoclassical collisionality is varied.

In the high-collisionality limit, as $\nu_i^*$ approaches unity and beyond, collisions scatter ions across the trapped-passing boundary on a bounce time or less, progressively destroying the mirror-confined orbit structure. In that limit the distribution at the midplane relaxes toward an approximately Maxwellian shape. \deleted{, so the trapped fraction no longer represents a dynamically isolated mirror-confined population. Instead, it approaches the geometric Maxwellian phase-space fraction above the loss-cone boundary, which for this field-line geometry is approximately 0.66. The excess trapped fraction above this Maxwellian value at low $\nu_i^*$ is therefore the more physically meaningful signature of kinetic mirror trapping and a trapped fraction of 0.66 signifies a more conventional fluid-like SOL.} The higher-recycling case has a ion and electron collisionality of 13.7 and 3.4. Since \gke is a full-$f$ solver it could be that the tail of the ion distribution function is moderately mirror trapped, but not the bulk, that might explain the modest poloidal potential the high-recycling case develops, fig~\ref{fig:resConvergence_kperp_phiDTe} and the modest trapped fraction seen in fig~\ref{fig:fTrap_nuStar}. However, at lower collisionality, a significant fraction of the distribution function can end up in the trapped region, for the low recycling base case, $\sim$ two thirds of the distribution function is in the trapped region, as seen in fig~\ref{fig:fTrap_nuStar}. 

\begin{figure}[htbp]
  \centering
  \includegraphics[width=0.85\textwidth]{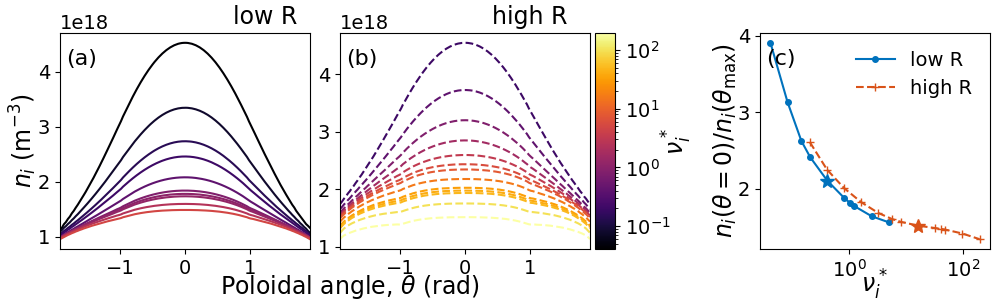}
  \caption{SOL plasma poloidal density profiles in simulations of the low-recycling case (a) and high-recycling case (b) with varying collisionality $\nu_i^*$, obtained by scaling the collision frequencies. (c) Poloidal density peaking as a function of collisionality measured as the ratio of the midplane to the sheath entrance ion density; cases with unscaled collision frequencies are indicated by the stars.}
  \label{fig:den_nuStar}
\end{figure}

The source parameters in the 1x2v simulations are adjusted quantities: the electron source temperature, ion source temperature, and particle fueling rate are chosen so that the simulations reproduce the measured or estimated midplane density and temperatures for each recycling state. Because these source parameters differ between the low-recycling and high-recycling cases, and because simulations are on a single field line neglecting cross field transport, direct comparison of end-plate particle fluxes has limited physical relevance. In steady state, the integrated heat and particle fluxes to the sheath must balance the imposed particle and power sources, so they mostly reflect the chosen source strength rather than the intrinsic parallel transport response. A more useful diagnostic is therefore the effective SOL residence time: how much particle or thermal energy content is maintained in the field-line domain for a given source rate.

\begin{figure}[htbp]
  \centering
  \begin{subfigure}[b]{0.45\textwidth}
    \includegraphics[width=\textwidth]{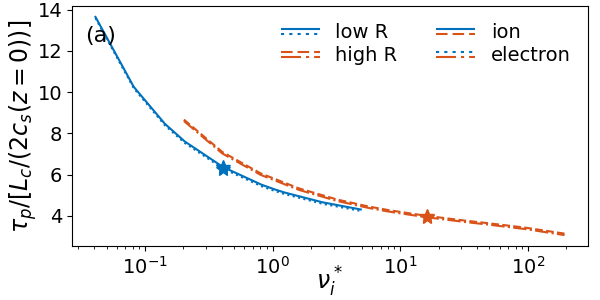}
  \end{subfigure}
  \begin{subfigure}[b]{0.45\textwidth}
    \includegraphics[width=\textwidth]{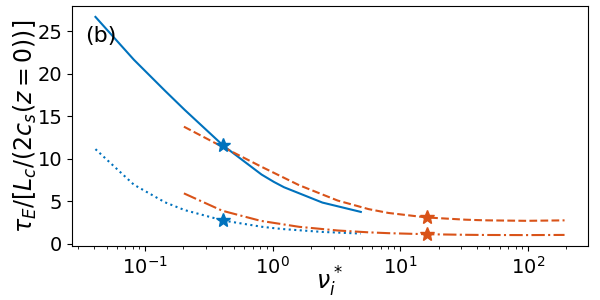}
  \end{subfigure}
  \caption{SOL particle (a) and energy (b) confinement times (normalized to the sound transit time) as a function of ion collisionality $\nu_i^*$ for simulations of the low-recycling (blue) and high-recycling (orange) cases, for ions (solid, dashed) and electrons (dotted, dash-dot).}
  \label{fig:tau_nuStar}
\end{figure}

Figure~\ref{fig:tau_nuStar} shows the SOL particle and energy confinement times inferred from this balance as a function of ion neoclassical collisionality. The particle confinement time is estimated from the ratio of the total number of particles in the SOL to the integrated particle source, while the energy confinement time is estimated from the ratio of the total thermal energy in the SOL to the imposed source power. Both are normalized to a simple sheath-limited transit time, $\tau_{\parallel 0}=L_z/(2c_s)$, where $c_s=\sqrt{(T_{e0}+5T_{i0}/3)/m_i}$ is evaluated using the midplane temperatures. This normalization removes the trivial change in parallel sound speed between the hotter low-recycling case and the colder high-recycling case, allowing the remaining variation to be interpreted as a change in effective parallel confinement.

The main result in figure~\ref{fig:tau_nuStar} is that the normalized SOL confinement times increase strongly as collisionality is reduced, especially for the low-recycling case. This is consistent with the interpretation developed above: at low $\nu_i^*$, mirror trapping and kinetic phase-space structure allow particles and energy to remain in the SOL longer than expected from a simple sheath-limited fluid transit estimate. At higher collisionality, collisions more efficiently destroy the mirror-confined orbit structure and push the system back toward the more conventional fluid-like regime. Thus figure~\ref{fig:tau_nuStar} connects the phase-space evidence for trapping to a macroscopic transport consequence: enhanced parallel particle and energy residence times along the modeled field line in the low-recycling, weakly collisional SOL. We hypothesize that this enhanced residence time may contribute to the observed poloidal density peaking in the low-recycling LTX-$\beta$ SOL. A quantitative connection to cross-field transport is therefore left for future work.

\added{Reduced upstream parallel electron heat flux relative to the local classical prediction has been obtained in kinetic SOL simulations without magnetic mirroring. In ALLA, preferential escape of higher-energy electrons depletes the upstream heat-carrying population~\cite{Batishchev1999}, while in KIPP, nonlocal electron transport and the asymmetric loss of electrons through the sheath modify the conductive heat flux~\cite{Zhao2017}. These effects can suppress the heat flux upstream and enhance it near the target relative to the local classical prediction. Both models include electrostatic regulation of electron losses. In the weakly collisional, low-recycling regime considered here, inclusion of the mirror force is expected to further inhibit parallel transport by retaining ions, with the ambipolar potential regulating electron losses to maintain quasineutrality. When the magnetic-field magnitude is held constant, removing the mirror force while retaining the same connection length and source parameters, the density profiles become substantially flatter, with peak densities approximately 3.7 and 2.7 times lower for the low and high-recycling cases, respectively. These comparisons support an additional contribution from magnetic trapping to plasma retention, beyond kinetic heat-flux suppression alone. Further details are given in \ref{sec:constBsims}.}

\section{Discussion and Conclusion} \label{sec:conclusion}

We have used gyrokinetic simulations with one configuration-space dimension and two velocity-space dimensions, using \gke to study the open-field-line SOL response associated with high and low-recycling \ltx conditions. The experimentally constrained low-recycling case is hotter and more weakly collisional than the high-recycling case, and develops a substantially larger electrostatic-potential variation along the modeled field line. The comparison with an adiabatic Boltzmann-electron closure of the field equations shows that the low-recycling density response cannot be recovered without kinetic electrons, indicating that the ambipolar potential and non-Maxwellian electron dynamics are important parts of the parallel force balance in this regime.

Collisionality scans further show that, as the ion neoclassical collisionality is reduced below unity, the ion distribution develops stronger mirror-trapped phase-space structure and the normalized particle and energy confinement times increase, most strongly for the low-recycling case. These results suggest that low-recycling lithium-wall operation in \ltx accesses a SOL regime in which weak collisionality, mirror trapping, and kinetic-electron effects increase the normalized parallel residence time relative to conventional high-recycling fluid expectations. We also document using experimental data an increase in the estimated SOL density fall-off length $\lambda_{n,\mathrm{SOL}}$ as the low-recycling SOL warms.

\added{The omission of neutral interactions and cross-field drifts is a limitation of the present field-line model. Across the relevant discharge ensembles, estimated SOL neutral densities of $1$--$5\times10^{16}\,\mathrm{m}^{-3}$ give charge-exchange lifetimes that increase from approximately $0.8$ ms at higher recycling to $4$ ms at low recycling, while curvature-induced guiding-center drift times are estimated at $1.6$--$3.9$ ms~\cite{maan2024recycling}. The low-recycling charge-exchange lifetime is therefore more than an order of magnitude longer than the approximately $0.17$ ms parallel particle residence time obtained for the low-recycling reference case. Together with the longer drift times, this supports the isolation of parallel kinetic dynamics as a first approximation. This ordering does not establish that all perpendicular transport is negligible: $E\times B$ transport and neutral interactions may affect the radial SOL structure and become more important at higher neutral densities or longer residence times.}

\added{Impurity interactions and radiation are also omitted. Impurity sources in \ltx depend strongly on lithium-conditioning history, with repeated lithium deposition associated with reduced impurity influx and improved recycling control compared with the initial transition from stainless-steel operation~\cite{maan_ppcf_2021,maan2026photodiode}. Relatively weak radiative losses are therefore expected for the well-conditioned, low-recycling discharge ensembles motivating the present study, consistent with their previous interpretation~\cite{maan2024recycling}. Nevertheless, radiation from the remaining Li, C and O impurities is not included in the present simulations and could modify the quantitative energy balance and parallel transport. Multistep atomic processes can substantially reduce electron heat loads in cold, dense divertor plasmas~\cite{Tskhakaya2023}; however, these effects are expected to be less important in the hot, low-density SOL considered here. Non-Maxwellian electron distributions can also modify impurity ionization and radiation rates, including those of lithium~\cite{Power2025}. The consequences for the energy balance depend on impurity concentration as well as the electron distribution. These radiation effects are therefore expected to be more important in high-recycling regimes, such as those in lithium rich vapor-box divertors, which deliberately use a dense lithium vapor for volumetric cooling~\cite{Goldston2016vaporbox}, than in the hot, low-density SOL sought by low-recycling flowing-lithium concepts such as divertorlets, which limit evaporation by removing heat through liquid flow~\cite{Fisher2020divertorlets}. Radiation would introduce an additional energy-loss channel and could modify collisionality, the ambipolar potential and particle residence times through changes in the electron distribution. The present results therefore characterize parallel transport without radiative losses, rather than the complete SOL energy balance. A recently developed kinetic line-radiation operator in \gke provides a route to assessing these effects self-consistently~\cite{Roeltgen2025}.}

The experimentally observed broadening of the SOL density profile can be considered along with a simple particle-balance argument. Particles that diffuse radially across the SOL are also lost along the open field line to material surfaces. If radial transport is represented by an effective cross-field diffusivity $D_\perp$, and parallel loss is represented by an effective field-line residence time $\tau_\parallel$, a broader radial density profile is expected when particles either diffuse outward more rapidly or remain on the field line longer before being lost. For an exponential SOL density profile, this balance gives

\begin{equation}
\lambda_n \sim \sqrt{D_\perp \, \tau_\parallel}.
\end{equation}

Using the $\lambda_{n,\mathrm{SOL}}$ values from the two profile-fit time points shown in figure~\ref{fig:lambdaNSOL_time} and the effective particle residence times from the simulations as shown in figure~\ref{fig:tau_nuStar} (90~$\mu$s and 168~$\mu$s), this scaling would imply an effective cross-field diffusivity that increases from $D_\perp \sim 3$ m$^2$/s when the gas puff is on and flooding the SOL with cold neutrals to $D_\perp \sim 25$ m$^2$/s after the gas puff terminates and the edge temperature increases. The observed broadening would require enhanced radial transport, increased parallel residence time, or some combination of the two. From the SOL particle-balance perspective, the puff-on phase behaves more like a cold, high-recycling boundary because of the neutral source, while after the puff terminates the SOL returns toward the warmer, lower-collisionality regime in which the simulations find longer normalized parallel residence times. The increase in the implied effective $D_\perp$ points to perpendicular transport physics not captured in the present 1D model.

\added{The stronger poloidal variation of the electrostatic potential also implies a radial $E\times B$ drift through its interaction with the toroidal magnetic field, providing an additional mechanism for redistribution between field lines. As parallel residence increases, perpendicular redistribution can become important before particles are lost to the limiters, modifying the profiles, trapped population, and loss balance. Comparable parallel-loss and perpendicular-transport times are already encountered in conventional SOL models~\cite{goldston2012,Russell2021}. The absence of this self-consistent coupling is a central limitation of the present study, particularly when extrapolating toward increasingly low collisionality.}

Since particle and energy losses are coupled in this low-recycling, weakly collisional regime, changes in the balance between parallel residence time and cross-field spreading will also affect the radial heat-flux width. This raises an important unanswered question about the nature of particle and energy transport in the low-recycling SOL\deleted{, and motivates future work that couples the kinetic parallel dynamics to perpendicular transport, neutral fueling, and wall pumping.}\added{. Self-consistent multidimensional kinetic simulations incorporating perpendicular drifts, neutral fueling, and wall pumping are therefore the highest-priority extension of this study. Together with experimental measurements of the SOL heat-flux width, these calculations are needed to establish how low-recycling operation modifies the distribution of heat loads on plasma-facing surfaces. We hope to explore some of these effects in future experiments, planned on ST40 with lithium walls~\cite{Asunta2026ST40}.}

\section*{Acknowledgements}

We thank Akash Shukla for discussing oscillations in the constant $B$ case with us. Support for E. Perez was provided by the Department of Energy (DOE) Science Undergraduate Laboratory Internships (SULI) program. Funding for M. Francisquez, A. Hakim and G. W. Hammett was provided by the CEDA project as part of DOE's Scientific Discovery through Advanced Computing (SciDAC) program, and the Distinguished Science Fellow program under contracts DE-AC02-09CH11466 and DE-FG02-04ER54742, as well as DOE’s Distinguished Scientist program and LDRD grants via DOE contract DE-AC02-09CH11466 for the Princeton Plasma Physics Laboratory. A. Maan, R. Majeski, and D. P. Boyle were supported by DE-AC02-09CH11466, and M. Dorf was supported by DE-AC52-07NA27344. This research used the Stellar cluster supported by Princeton Research Computing, a consortium of groups including the Princeton Institute for Computational Science and Engineering (PICSciE) and the Office of Information Technology’s High Performance Computing Center and Visualization Laboratory at Princeton University. The United States Government retains a non-exclusive, paid-up, irrevocable, world-wide license to publish or reproduce the published form of this manuscript, or allow others to do so, for United States Government purposes. 
\section*{Data Availability}
The raw data is available upon request. All input files to reproduce it can be found in \url{https://github.com/ammarhakim/gkyl-paper-inp/tree/master/2026_NF_LTX_1x2v}. Other data used in the paper is hosted by Princeton Data Commons \cite{gke_data}.

\appendix

\section{Getting \gke~and reproducing results} \label{sec:getGkeyll}

Readers may reproduce our results and also use \gke~for their applications. The code and input files used here are available online.
Full installation instructions for \gke~are provided on the \gke~website (\url{https://gkeyll.readthedocs.io}). The input files used here are under version control and can be obtained from the repository at \url{https://github.com/ammarhakim/gkyl-paper-inp/tree/master/2026_NF_LTX_1x2v}.

\section{Convergence studies} \label{sec:convergence}

In this appendix we show data from convergence studies done to ensure that our findings are obtained with a suitable choice for final simulation time, number of cells, velocity grid extents, and $\kperp\rho_s$ in kinetic electron simulations.

\begin{figure}[h]
  \centering
  \includegraphics[width=0.65\textwidth]{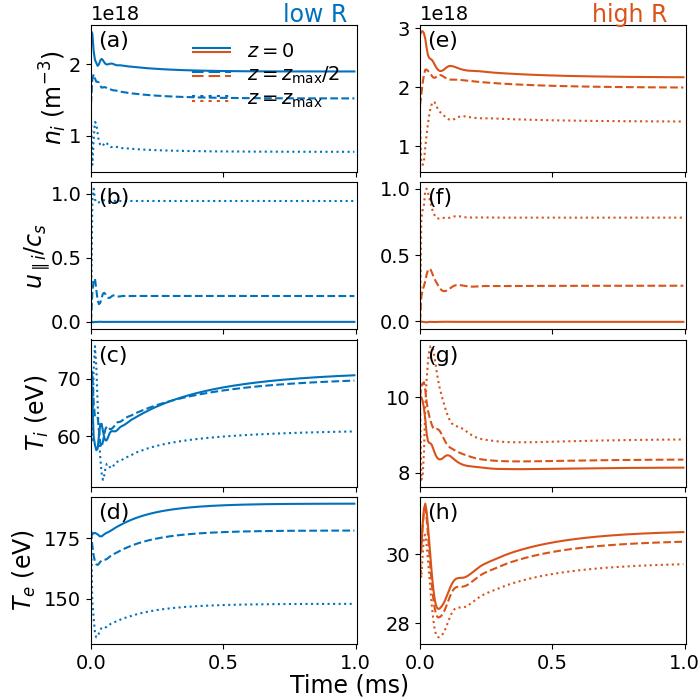}
  \caption{Time traces of the moments in simulations of low R (left column) and high R (right column) shots at three different locations: outboard midplane (solid), half-way to the sheath entrance (dashed) and sheath entrance (dotted).}
  \label{fig:nFluxTT}
\end{figure}

In figure~\ref{fig:nFluxTT} we present time traces of the ion density, parallel drift speed and temperature, as well as the electron temperature at three key locations: the midplane ($z=0$), half way between the midplane and the sheath entrance ($z=z_{\mathrm{max}}/2$), and the sheath entrance ($z=z_{\mathrm{max}}$). We see that, for both the low and high \recR~simulations, these quantities are approximately at steady state by the time the simulation is terminated, at $t=1$ ms. Although not shown here, we also tried running these much longer ($t>3$ ms), but the results and analysis do not change. This also applies to the simulations with oscillations presented in~\ref{sec:constBsims}; specifically, running the low \recR~simulation with $B=B_0$ for much longer times does not make oscillations go away.

\begin{figure}[h]
  \centering
  \includegraphics[width=0.65\textwidth]{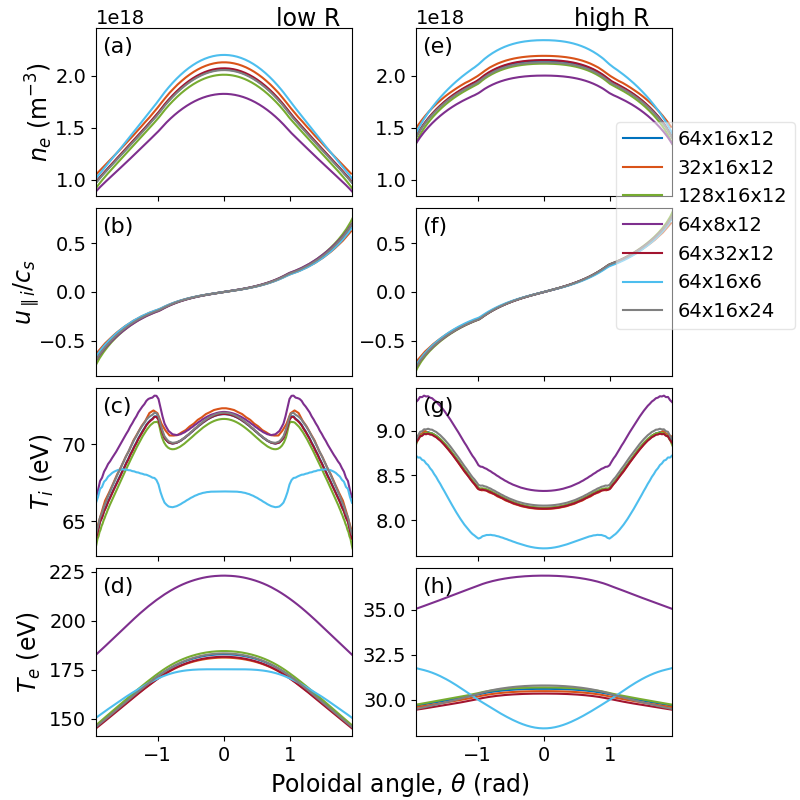}
  \caption{Ion number density (a,e), parallel speed (b,f) and temperature (c,g), and electron temperature (d,h), in simulations of low R (left column) and high R (right column) shots, using kinetic electrons and resolutions doubling or halving the number of cells in each direction.}
  \label{fig:resConvergence_Te}
\end{figure}

We also ensured that the results presented in the body of this manuscript are representative of what would be obtained at much higher resolution. From the resolution convergence study presented in figure~\ref{fig:resConvergence_Te} we draw the following conclusions. Obtaining reasonably converged $T_{\perp i}$ (and thus $T_i$) requires at least 12 cells along $\mu$. Likewise, producing a $T_e$ close to that obtained at the highest $\vpar$ resolution requires $\Nvpar\geq16$. Lastly, according to figure~\ref{fig:resConvergence_Te}(a) and (e), the density is reasonably converged with 64 cells in $z$. Hence, the resolution chosen for the reference simulations in section~\ref{sec:baseSims} is $64\times16\times12$ cells.

\begin{figure}[h]
  \centering
  \includegraphics[width=0.65\textwidth]{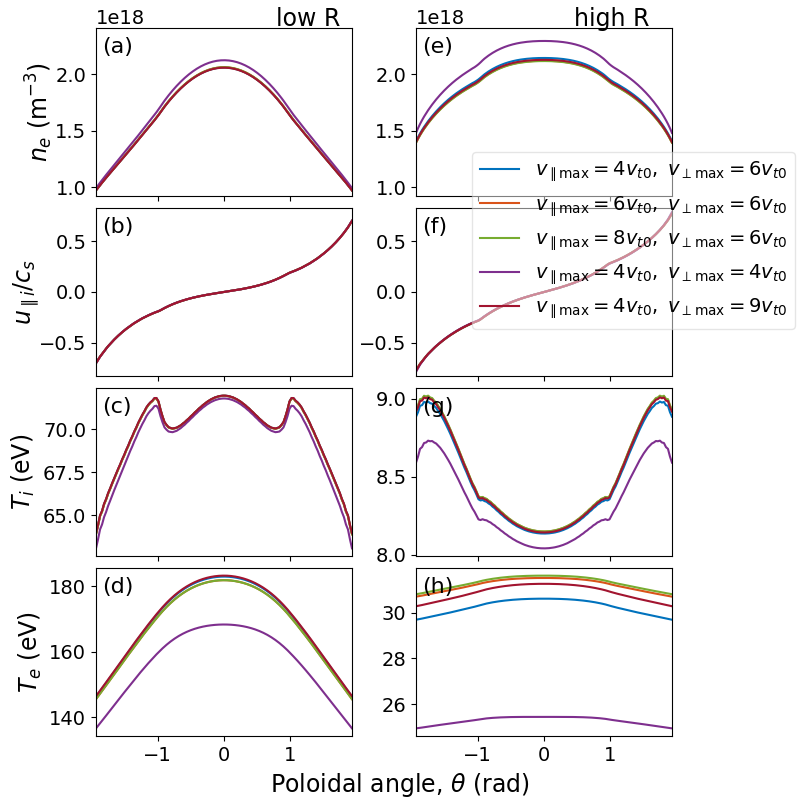}
  \caption{Ion number density (a,e), parallel speed (b,f) and temperature (c,g), and electron temperature (d,h), in simulations of low R (left column) and high R (right column) shots, using kinetic electrons and varying the velocity extents in $\vpar$ and $\mu\propto\vperp^2$.}
  \label{fig:vmaxConvergence_moms}
\end{figure}

A similar analysis was performed for the extents of the velocity-space grid. In figure~\ref{fig:vmaxConvergence_moms} we show that changing the velocity space limits so that the maximum $\vpar$ or $v_\perp$ is between $4v_{ts0}$ and $9v_{ts0}$ changes the results relatively little. The biggest change is seen when the $v_{\perp,\mathrm{max}}$ is too low, as it is for the purple curve. If $v_{\perp,\mathrm{max}}$ is too low, the densities come out too high, and the temperatures too low. By $v_{\perp,\mathrm{max}}=6v_{ts0}$ the simulation is very close to converged.

\begin{figure}[h]
  \centering
  \includegraphics[width=0.65\textwidth]{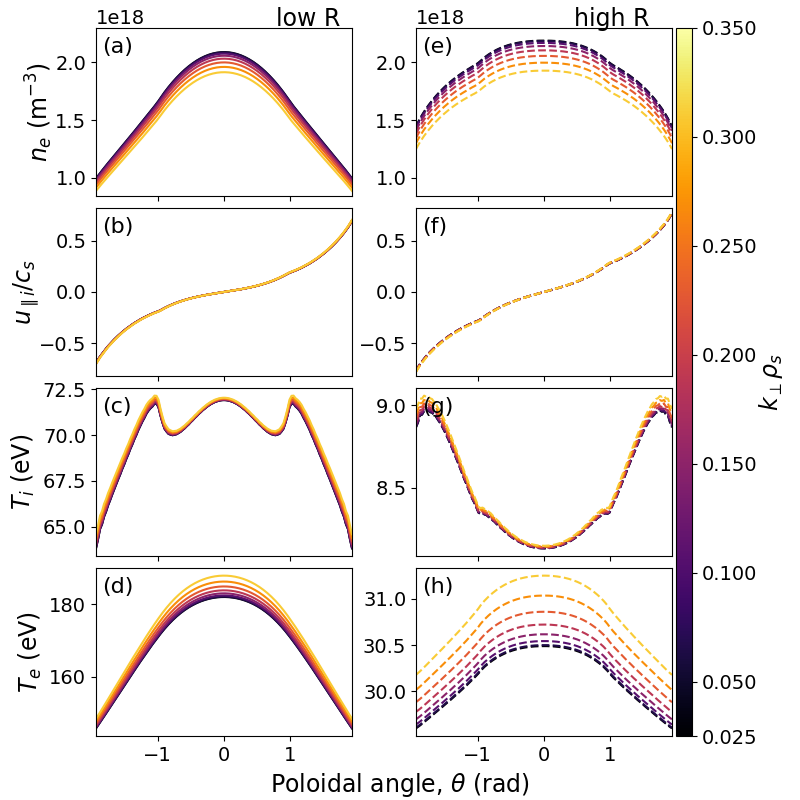}
  \caption{Moments in simulations of low R (left column) and high R (right column) cases as a function of the $\kperp\rhos$ parameter.}
  \label{fig:resConvergence_kperp}
\end{figure}

Lastly, we show what happens if smaller and larger values of the $\kperp$ parameter in the kinetic electron field model (see equation~\ref{eq:poisson1d}) are used. Figure~\ref{fig:resConvergence_kperp} presents the final moments of ions and electrons for both low and high \recR~simulations as a function of the $\kperp$ parameter. In general the moments are weakly sensitive to $\kperp\rho_s$, provided that a sufficiently low value is chosen. The reference simulations in section~\ref{sec:baseSims} used $\kperp\rho_s=0.15$, for which the moment profiles are relatively close to their asymptotic value at $\kperp\rho_s=0.025$. Only the electron temperature at high \recR~(figure~\ref{fig:resConvergence_kperp}(h)) appears to be more sensitive to this free parameter, but note that there is a suppressed zero in that plot, causing the differences to appear larger than they really are.

\section{Simulations without mirror force} \label{sec:constBsims}

The mirror force ($\propto\partial_z(\mu B)$) plays an important role in the formation of edge profiles in \ltx. The low collisionality (due to high temperatures and low densities) and considerable mirror ratio ($\sim$ 4) give the mirror force term in equation~\ref{eq:gkeq1d} greater importance than in the edge of high-recycling scenarios. This effect is typically excluded from edge fluid simulations with codes such as UEDGE~\cite{Rognlien1992}, SOLPS~\cite{Wiesen2015}, EDGE2D~\cite{Simonini1994}, SOLEDGE~\cite{Bufferand2021}, BOUT++~\cite{Dudson2015}, GBS~\cite{Halpern2016}, GRILLIX~\cite{Stegmeir2018} and TOKAM3X~\cite{Tamain2016}. Using fluid models that neglect the mirror force would produce very different characterizations or predictions of the SOL in these low collisionality, low-recycling regimes. In order to demonstrate what would happen were one to neglect the mirror force, in this appendix we show simulations with a constant magnetic field amplitude $B=B_0=\mathrm{const}$. \added{Unless stated otherwise, the} connection length, the source parameters, and other inputs are kept the same as they were for the $B=B(z)$ simulations.

\begin{figure}[htbp]
  \centering
  \begin{subfigure}[b]{0.45\textwidth}
    \includegraphics[width=\textwidth]{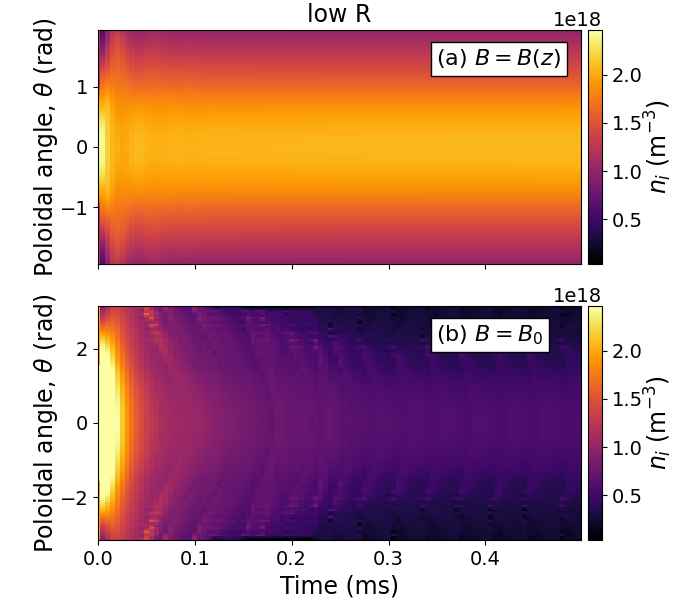}
  \end{subfigure}
  \begin{subfigure}[b]{0.45\textwidth}
    \includegraphics[width=\textwidth]{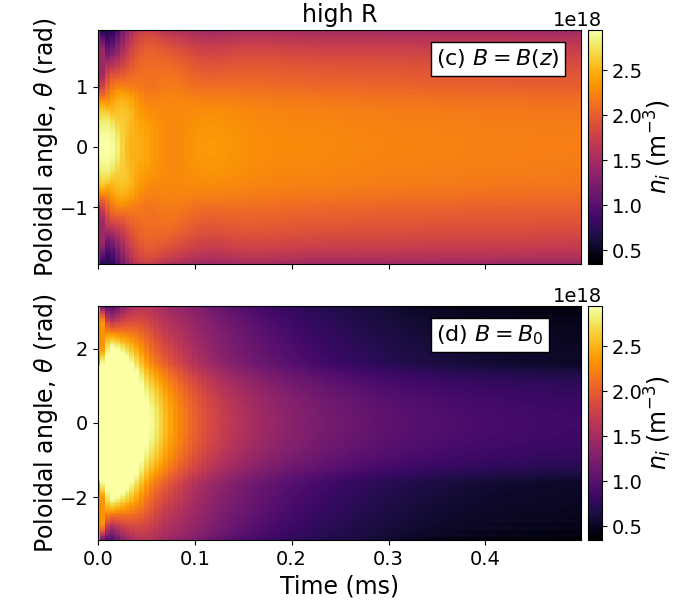}
  \end{subfigure}
  \caption{Ion density as a function of space and time for low (left) and high (right) \recR~simulations, and for $B=B(z)$ (top) and $B=B_0$ (bottom).}
  \label{fig:constB_ni_vs_t}
\end{figure}

Two major changes are observed when the magnetic field amplitude is set to a constant: the density profiles are less peaked, and oscillations arise near the sheath entrance \added{in} the low-recycling simulation. Both of these can be appreciated in the space-time evolution of the ion density given in figure~\ref{fig:constB_ni_vs_t}. The steady state peak ion density ($n_i(\theta=0,t\approx1\mathrm{ms})$) is much lower when using constant magnetic fields for both low and high $R$. Additionally, in figure~\ref{fig:constB_ni_vs_t}(b) we can appreciate oscillations that persist for $t>0.3$ ms.

\begin{figure}[htbp]
  \centering
  \includegraphics[width=0.65\textwidth]{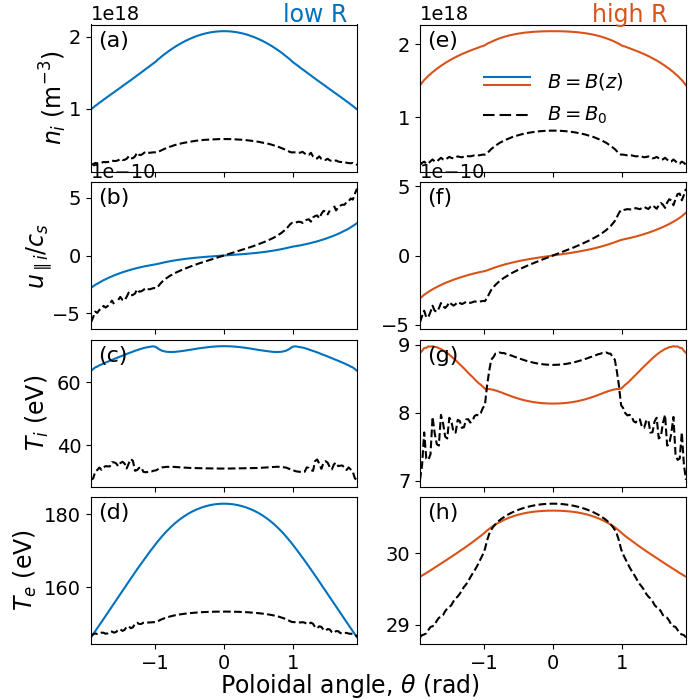}
  \caption{(a) Ion number density, (b) parallel speed, (c) ion temperature, and (d) electron temperature for the low-recycling case averaged over $t\in\left[0.8~\mathrm{ms},~1~\mathrm{ms}\right]$. Panels (e)--(h) show the same quantities for the high-recycling case. The $B=B(z)$ results are shown in solid lines, while the $B=B_0$ results are shown in dashed lines.}
  \label{fig:nTprofiles_constB}
\end{figure}

A density flattening is more clearly appreciated in the time-averaged, steady state profiles given in figure~\ref{fig:nTprofiles_constB}(a) and (e). We see that the constant $B$ equilibrium profiles have peak densities lower by factors of approximately 3.7 and 2.7 for the low and high \recR~scenarios, respectively, compared to their inhomogeneous $B$ counterparts. The stronger flattening at low \recR~is attributed to the loss of trapped (banana) particle orbits, which is particularly important for low \recR, as exemplified by the larger trapped particle fraction quantified in figure~\ref{fig:fTrap_nuStar}. The lack of mirror trapping is also evident in the \deletedRtwo{vanishing}\addedRtwo{reduced} off-midplane peaks in $T_i$, which are \deletedRtwo{caused by}\addedRtwo{associated with} the increase in $T_{\perp i}$ as a result of conservation of $\mu$ in a varying magnetic field. \addedRtwo{Since temperature is a local moment of the distribution function, this mirror-induced redistribution and the spatial variation of the contributing particle populations can modify the temperature profile without violating single-particle energy conservation.} Such peaks do not exist in $T_e$ because\added{, although subject to the same $\mu$-conservation that causes $T_i$ to increase off mid-plane, the electrons have much higher energy (180 eV) at the midplane and manage to escape mirror trapping and transport their energy out of the domain, thus causing the temperature to be peaked in the center.}

As mentioned in this section, the low \recR~simulation with $B=B_0$ develops oscillations near the boundaries. These oscillations propagate both downstream and upstream, and produce quasiperiodic patterns in the potential, the moments and the distribution function. \added{We performed a very extensive study of these oscillations (over several years) in order to determine the origin of these oscillations or find what algorithms or physics could stabilize them. We experimented with the following:}
\begin{itemize}
    \item \added{Decreasing time step $\Delta t$.}
    \item \added{Considerably increasing the resolution in all dimensions.}
    \item \added{Changing the extents of the grid in $\vpar$ and $\mu$.}
    \item \added{Altering the initial conditions (ICs); e.g. we tried a Gaussian $n_s(t=0)$ instead of figure~\ref{fig:ic_mom}(a), or non-Maxwellian ICs.}
    \item \added{Modifying the source shape or strength.}
    \item \added{Increasing or decreasing the connection length.}
    \item \added{Scanning the electron-ion temperature ratio (oscillations are stabilized for small $T_{e0}/T_{i0}$).}
    \item \added{Scanning the collision frequency (oscillations are stabilized at large $\nu_{ss}$).}
    \item \added{Several algorithms intended to enforce the Bohm sheath criterion on $u_{\parallel i}$.}
    \item \added{Different ways of constructing the numerical flux and discretizing the geometric factors arising in Gkeyll's field-aligned coordinate system.}
\end{itemize}
\added{None of the above produced a convincing story that would portray these oscillations as either numerical or physical.}

\begin{figure}[htbp]
  \centering
  \includegraphics[width=0.65\textwidth]{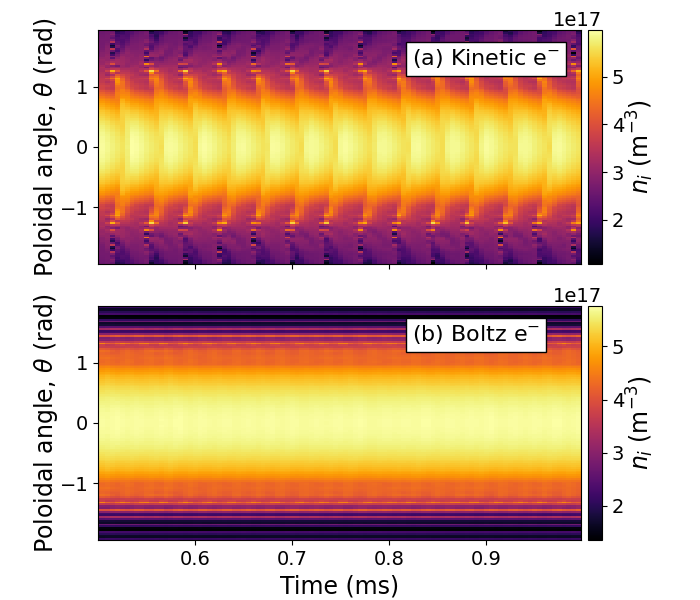}
  \caption{Ion density in low \recR~simulations with $B=B_0$ and kinetic (a) or Boltzmann (b) electrons, as a function of time and space. The first 0.5 ms are omitted to avoid the contrast caused by transients that ICs induce.}
  \label{fig:ni_vs_t_constB}
\end{figure}

\added{In order to simplify the system and attempt to understand the origin of these oscillations in the constant $B$ case, we switched to using the Boltzmann electron model (see section~\ref{sec:boltz}). The Boltzmann electron model also removes the conducting sheath BC~\cite{Shi2017thesis} in favor of an absorbing BC for ions. Even with Boltzmann electrons, the oscillations emerge at the same location (figure~\ref{fig:ni_vs_t_constB}(b)), although they have a different amplitude and are relatively constant in time.} 

\begin{figure}[htbp]
  \centering
  \includegraphics[width=0.65\textwidth]{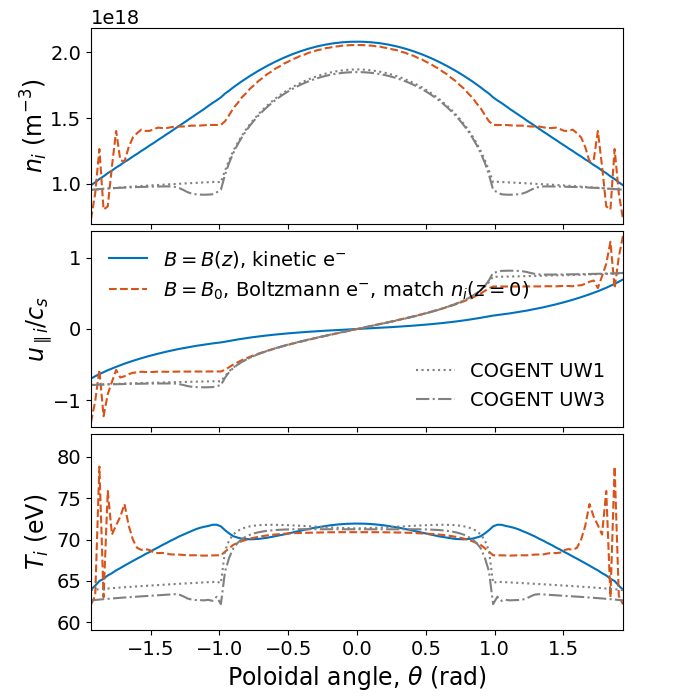}
  \caption{For \gke's standard algorithm, the ion density \deleted{(a)} \added{(top)}, parallel drift speed normalized to $c_s=\sqrt{(\Tpare+3\Tpari)/m_i}$ \deleted{(b)} \added{(middle)} and temperature \deleted{(c)} \added{(bottom)} in low \recR~simulations with $B=B(z)$ and kinetic electrons (solid blue), and $B=B_0$ and Boltzmann electrons after adjusting the sources so that $n_i(z=0)$ roughly matches the peak density of the case with inhomogeneous $B$ and kinetic electrons (dashed orange). The blue line consists of the final steady state, while the orange line is the result of a 1 ms time average.}
  \label{fig:moms_kinVSboltz_match}
\end{figure}

Notice from figure~\ref{fig:nTprofiles_constB} that the low \recR~simulation with $B=B_0$ arrives at equilibrium profiles with very different values. This observation begs the question of whether the oscillations are produced by an instability that only arises in this very different parameter regime. In order to rule out that possibility, we adjusted the density source rate and the source temperature so that the $B=B_0$, Boltzmann electron simulation culminates in profiles in the same range as those in the $B=B(z)$, kinetic electron simulation. The result is given in figure~\ref{fig:moms_kinVSboltz_match}. We see now that despite having a density and temperature close to the more realistic simulation, the $B=B_0$ simulation with Boltzmann electrons still develops oscillations near the boundaries. For reference we also ran this case with the COGENT gyrokinetic code~\cite{Dorf2016}, essentially solving the same equation with the same setup. The result is also shown in \added{the top panel of} figure~\ref{fig:moms_kinVSboltz_match}, where we see that for both the first-order (UW1) and third-order (UW3) upwind discretizations \added{COGENT did not develop oscillations near the boundaries. This suggested that either COGENT had enough numerical dissipation (e.g. from upwinding) to suppress these (possibly physical) oscillations, or that they are a numerical artifact produced by Gkeyll's numerics.}

\added{For lack of time to dig into COGENT's numerics we explored if Gkeyll's algorithm could be causing these oscillations. The short answer appears to be yes. Explaining how Gkeyll's numerics can cause such oscillations, and how to remedy them, in detail will have to be postponed to a future manuscript, but we provide a high-level description here, and a demonstration that an alternative algorithm is more stable. In simple terms, the 1D1V limit of the Cartesian, constant $B$ gyrokinetic equation (neglecting collisions and sources) reads:}
\begin{equation} \label{eq:dg1}
    \added{\pd{f_s}{t} + \pd{^{\uparrow\vpar}}{z}\left(\vpar f_s\right) + \pd{^{\uparrow a_{\parallel}}}{\vpar}\left(a_{\parallel} f_s\right) = 0, \qquad a_\parallel = -\frac{q_s}{m_s}\pd{\phi}{z}.}
\end{equation}
\added{In this equation we have added a superscript to the derivatives to symbolize that these are upwinded derivatives. A weak upwinded derivative of a scalar $A$, for example $\partial_z^{\uparrow\vpar}A$, follows from projecting this derivative onto the $k$-th finite element or DG basis $\psi^{k}_j$ in cell $j$, using integration by parts}
\begin{equation}
    \added{\int_{z_{j-1/2}}^{z^{j+1/2}}\mathrm{d}z\,\psi^{k}_j\,\partial_z^{\uparrow\vpar}A = \psi^{k}_j\,\widehat{A}\Big|_{z_{j-1/2}}^{z^{j+1/2}} - \int_{z_{j-1/2}}^{z^{j+1/2}}\mathrm{d}z\,A\,\partial_z\psi^{k}_j,}
\end{equation}
\added{and the numerically upwinded $A$:}
\begin{equation}
    \added{\widehat{A}_{j-1/2} = \frac{1}{2}\left(A^-_{j-1/2}+A^+_{j-1/2}\right)-\frac{\mathrm{sign}\left(\vpar\right)}{2}\left(A^+_{j-1/2}-A^-_{j-1/2}\right),}
\end{equation}
\added{where $-$/$+$ superscripts indicate the left/right value at a cell boundary. The $\vpar$ derivative in equation~\ref{eq:dg1} is upwinded similarly, using the sign of $a_\parallel$. The $z$ derivative in $a_{\parallel}\propto\partial_z\phi$ is not upwinded; it is simply computed locally from the intracell high-order representation of $\phi$. This is, in a nutshell, the algorithm in Gkeyll. The problem is that if one works out the discrete dispersion relation for this system, assuming that $\phi$ is computed as a continuous finite element scalar~\cite{Francisquez_2025}, an oscillatory mode can arise for subsonic parameter regimes.}

\begin{figure}[htbp]
  \centering
  \includegraphics[width=0.65\textwidth]{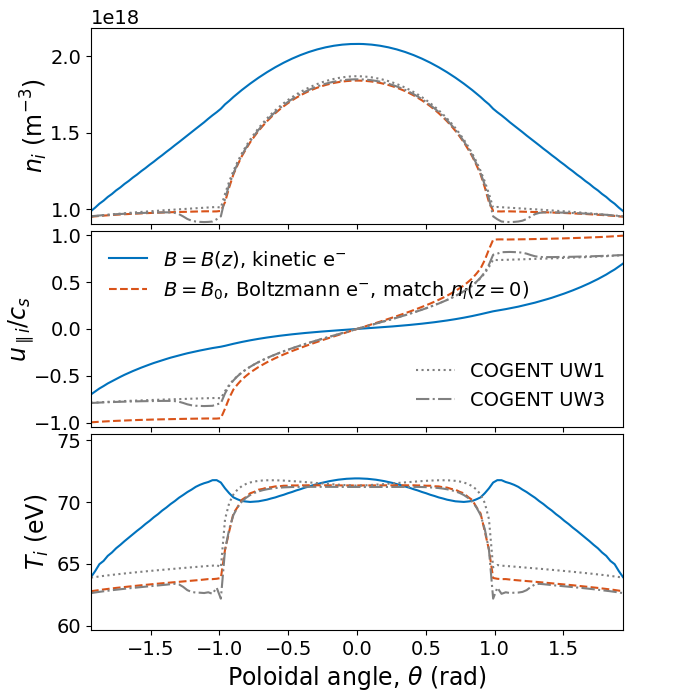}
  \caption{\added{For \gke's new matched adjoint algorithm, the ion density (top), parallel drift speed normalized to $c_s=\sqrt{(\Tpare+3\Tpari)/m_i}$ (middle) and temperature (bottom) in low \recR~simulations with $B=B(z)$ and kinetic electrons (solid blue), and $B=B_0$ and Boltzmann electrons after adjusting the sources so that $n_i(z=0)$ roughly matches the peak density of the case with inhomogeneous $B$ and kinetic electrons (dashed orange). The blue line consists of the final steady state, while the orange line is the result of a 1 ms time average.}}  \label{fig:moms_kinVSboltz_match_dgH}
\end{figure}

\added{One way to avoid this (we tried a couple) is to ensure that the derivative used to compute the electric field is the matched adjoint of the derivative in the position-space advection term. That is, to define a downwinded derivative $\partial_z^{\downarrow\vpar}\left(\cdot\right)$ such that}
\begin{equation}
    \added{\int\mathrm{d}z\,\mathrm{d}\vpar\,\partial_z^{\uparrow\vpar}A = - \int\mathrm{d}z\,\mathrm{d}\vpar\,\partial_z^{\downarrow\vpar}A.}
\end{equation}
\added{More explicitly, by computing the electric field using}
\begin{equation}
\added{\begin{aligned}
    \int_{z_{j-1/2}}^{z^{j+1/2}}\mathrm{d}z\,\psi^{k}_j\,\partial_z^{\downarrow\vpar}\phi &= \psi^{k}_j\,\check{\phi}\Big|_{z_{j-1/2}}^{z^{j+1/2}} - \int_{z_{j-1/2}}^{z^{j+1/2}}\mathrm{d}z\,\phi\,\partial_z\psi^{k}_j, \\
    \check{\phi}_{j-1/2} &= \frac{1}{2}\left(\phi^-_{j-1/2}+\phi^+_{j-1/2}\right)+\frac{\mathrm{sign}\left(\vpar\right)}{2}\left(\phi^+_{j-1/2}-\phi^-_{j-1/2}\right).
\end{aligned}}
\end{equation}
\added{For this to have an impact, $\phi$ cannot be represented as a continuous finite element quantity (as has historically been done in Gkeyll). We implemented this new algorithm in Gkeyll and found that the oscillations are indeed stabilized. For example, in figure~\ref{fig:moms_kinVSboltz_match_dgH} we once again show the ion moments for the $B(z)$ kinetic electron and $B=$const. Boltzmann electron Gkeyll simulations using this new matched adjoint algorithm, compared to the COGENT result. The Gkeyll and COGENT results now lie very close to each other. Additional tests of this new algorithm, with kinetic electrons and in higher dimensions, are deferred to a separate manuscript.}

\FloatBarrier
\section*{References}

\begingroup
\interlinepenalty=10000
\bibliographystyle{iopart-num}
\bibliography{main}
\endgroup

\end{document}